\documentclass{article} 
\usepackage{iclr2027_conference,times}
\usepackage{float}

\usepackage{amsmath,amsfonts,bm}

\def\eqref#1{equation~\ref{#1}}

\def\1{\bm{1}}

\DeclareMathAlphabet{\mathsfit}{\encodingdefault}{\sfdefault}{m}{sl}
\SetMathAlphabet{\mathsfit}{bold}{\encodingdefault}{\sfdefault}{bx}{n}

\usepackage{xurl}
\usepackage{multirow}
\usepackage{graphicx}
\usepackage{hyperref}
\usepackage{enumitem}
\usepackage{url}
\usepackage{wrapfig}
\newcommand{\lz}[1]{{\color{black}#1}}
\definecolor{cxypink}{HTML}{5A5A8B}
\newcommand{\cxy}[1]{{\color{black}#1}}

\usepackage{booktabs}
\usepackage{colortbl}
\usepackage{multirow}
\usepackage{caption}
\usepackage{subcaption}
\usepackage{placeins}
\usepackage{afterpage}

\title{\lz{RobotEQ 3.0: Towards Personalized Social Proactive Intelligence in Embodied Agents}}

\author{Shufan Zhang$^{1}$, Xinyi Che$^{1}$, Kuofei Fang$^{1}$, Xuehao Wang$^{1}$, Liyi Liu$^{1}$, Junqing Wu$^{1}$,\\
\textbf{Jiayi Cao$^{1}$, Ziyanghui Wang$^{1}$, Yanhan Huang$^{1}$, Chuyu Wu$^{1}$, Zheng Lian$^{1}$\thanks{Corresponding Author}\;} \\
{$^1$State Key Laboratory of Autonomous Intelligent Unmanned Systems, Tongji University}
}

\usepackage{float}
\usepackage{longtable}
\usepackage{array}
\usepackage{amsmath}
\usepackage{caption}
\usepackage{etoc}

\usepackage[most]{tcolorbox}
\usepackage{xcolor}

\definecolor{promptgray}{RGB}{112,112,112}
\definecolor{promptborder}{RGB}{180,180,180}

\newtcolorbox{promptbox}[1]{
  enhanced jigsaw,
  breakable,
  colback=white,
  colframe=promptborder,
  colbacktitle=promptgray,
  coltitle=white,
  title={Prompt: #1},
  title after break={Prompt: #1 (continued)},
  fonttitle=\bfseries\footnotesize,
  fontupper=\footnotesize,
  boxrule=0.5pt,
  titlerule=0pt,
  arc=0pt,
  outer arc=0pt,
  left=7pt,
  right=7pt,
  top=5pt,
  bottom=5pt,
  toptitle=1.5pt,
  bottomtitle=1.5pt,
  lefttitle=5pt,
  righttitle=5pt,
  before skip=6pt,
  after skip=9pt,
  before upper={\setlength{\parindent}{0pt}\setlength{\parskip}{3pt}\raggedright}
}

\iclrfinalcopy 
\begin{document}

\maketitle
\fancyhead{}
\renewcommand{\headrulewidth}{0pt}

\begin{abstract}
\lz{

Social Proactive Intelligence (SPI) is an emerging research area, aiming to shift embodied agents from reactive assistance toward proactively understanding human needs and executing socially desirable actions. Prior work has largely centered on the \emph{average user}. However, human expectations are inherently diverse, and prior work overlooks \emph{individual nuances}. To bridge this gap, we introduce \textbf{RobotEQ 3.0}, a benchmark for Personalized SPI. \textbf{(Dataset)} We first profile participants via a structured questionnaire covering factors that are correlated with human expectations of embodied agents, such as basic demographics and personality traits. Participants then select their preferred actions from a set of candidates. Unlike prior SPI benchmarks that focus on assessing behavioral appropriateness, our task centers on predicting the actions preferred by a specific user, thereby capturing human subjectivity. The resulting dataset establishes explicit links between individual traits and behavioral preferences. \textbf{(Solution)} We observe substantial inter-annotator variance, confirming that user preferences over actions are highly individualized. This motivates our exploration of Personalized SPI, in which user traits serve as additional inputs to predict individual preferences. Experimental results show that incorporating user traits can aid personalized prediction. \emph{This work aims to shift the research paradigm from developing agents suited for the average user to designing systems tailored to specific individuals.}

}
\end{abstract}
\section{Introduction}

\lz{

Embodied AI has emerged as a promising research direction in both academia and industry, driven by its vast potential applications. Based on the reliance on explicit user commands, current research can be broadly categorized into \emph{reactive assistance} and \emph{proactive assistance}. The former relies on explicit user instructions during task execution, while the latter actively infers human needs and executes corresponding actions. As embodied agents become increasingly present in social environments, passively executing instructions is insufficient for real-world human-robot interaction~\citep{ding2024atombot,hou2024egosocialarena,munje2025socialnavsub}. Proactive assistance, by contrast, enables more user-friendly support, making it a more promising paradigm. Recently, \textbf{Social Proactive Intelligence} (SPI)~\citep{roboteq2026,che2026roboteqvideo} has further expanded this concept by incorporating social desirability and extending to more diverse and complex scenarios.

}

\lz{

SPI is currently in its early stages. RobotEQ \citep{roboteq2026} introduced the concept of SPI and characterized its key properties, aiming to realize socially desirable embodied agents in open-domain environments. RobotEQ-Video \citep{che2026roboteqvideo} extends the SPI research from static images to dynamic videos and ensures more comprehensive coverage of diverse contextual factors in benchmarking. However, prior works center on the \emph{average user}, overlooking that human preferences vary across \emph{individuals}. For instance, extroverted individuals may prefer more proactive assistance, while introverted individuals may favor quieter support. In real-world scenarios, multiple behaviors can be equally valid, yet users prefer different actions due to variations in personality and individual preferences~\citep{yang2024personalization,wang2024pbarl}. Capturing such user-specific differences is crucial for realizing personalized embodied agents~\citep{ozbey2026stars,chen2026knowubench}.

}

\begin{figure}[t]
\begin{center}
\includegraphics[width=\linewidth]{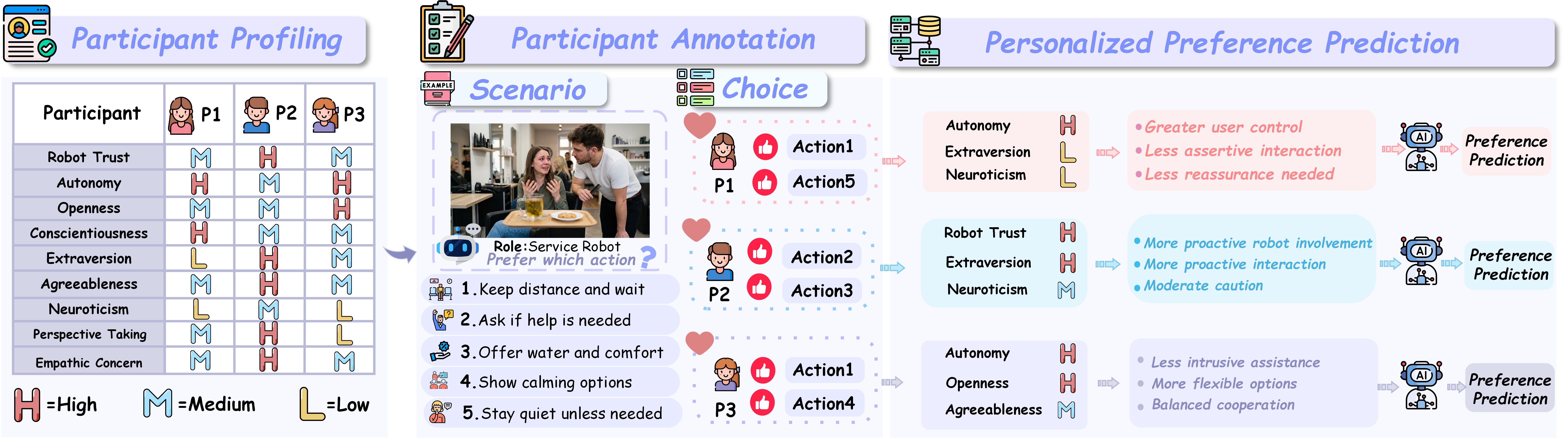}
\end{center}
\caption{\lz{\textbf{Overview of RobotEQ 3.0.} This paper collects participant profiles via a structured questionnaire, records each participant's preferred actions of embodied agents for each scenario, and investigates whether these profiles aid in predicting individual choices.}}
\label{fig:roboteq3_overview}
\end{figure}

\lz{

To address this limitation, we introduce \textbf{RobotEQ 3.0}, the first SPI benchmark that explicitly links human profiles to their action preferences for embodied agents. Departing from traditional works that rely on the \emph{average user}, this paper shifts the research focus to \emph{each individual}. To this end, we construct a dedicated dataset and explore solutions for personalized SPI. \textbf{(Dataset)} We first ask participants to complete a 78-item questionnaire related to human expectations of embodied agents, covering basic user information, robot experience, robot trust, personality, autonomy, and empathy. From this, we obtain structured profiles for each participant. We then present each participant with a set of reasonable candidate actions per scenario and ask them to select their preferred actions. This yields a dataset of 10 participants across 1,800 scenarios, totaling 18K participant-scenario annotations. \textbf{(Solution)} Beyond dataset construction, we benchmark 18 representative models to evaluate the impact of user profiles on individual preference prediction. Experimental results demonstrate that incorporating user profiles improves the performance on personalized SPI. To further enhance profile utilization, we propose an \emph{Iterative Profile-Aware Experience Induction} method, which summarizes human experience and dynamically retrieves relevant knowledge for personalized SPI. Thus, this paper aims to advance SPI research from majority-vote preferences to individualized ones. Figure~\ref{fig:roboteq3_overview} illustrates the overall pipeline of RobotEQ 3.0, spanning participant profiling, participant annotation, and personalized preference prediction. Our main contributions are summarized as follows:

}


\lz{

\begin{itemize}[leftmargin=2em]

    \item \textbf{(Task)} This paper proposes Personalized SPI, extending prior work by shifting the focus from population-level modeling to individual-level analysis.

    \item \textbf{(Dataset)} We introduce a dataset comprising rich individual profiles and corresponding user-preferred actions for embodied agents, providing a valuable resource for personalized SPI.

    \item \textbf{(Solution)} We study the efficacy of individual traits in personalized SPI and propose methods to better leverage these traits, offering insights for the development of personalized SPI systems.

\end{itemize}

}

\section{Related Work}

\lz{

\subsection{Social Proactive Intelligence}
\label{sec:SPI}

SPI has emerged as an important research topic in embodied AI. Mirroring the evolution of \emph{human-human interaction}, which has gradually shifted from functional utility to sociality, current \emph{human-robot interaction} research still centers on functional capabilities, leaving social properness largely underexplored. SPI aims to enable embodied agents to adhere to social expectations across diverse scenarios. \cite{fang2026roboteq} first introduced the concept of SPI, exploring population-level expectations of robot behaviors. \cite{che2026roboteqvideo} further scaled the dataset to multimodal inputs and introduced a fine-grained taxonomy to construct diverse embodied scenarios. However, existing works primarily rely on majority voting to produce unified labels, capturing only the consensus expectations of the general population. This approach overlooks the fact that human preferences for embodied agents are deeply tied to individual traits. Different people may favor different actions in the same scenario. Therefore, this paper extends SPI from population-level judgments to individual-level analysis, and introduces a new task \textbf{Personalized SPI}.

}

\lz{

\subsection{Personal Characteristic Modeling}
\label{sec:personal_characteristic}

Modeling personal profiles provides a well-established framework for explaining and predicting individual variations in decision-making processes \citep{yang2024personalization, wang2024pbarl}. Among the diverse factors within personal profiles, \emph{personality} is most strongly linked to human decision-making, and the Big Five model offers a widely validated framework for quantifying personality differences \citep{john1991bfi,kabacinska2025personality}. Beyond personality, \emph{basic demographics} such as age and gender are also widely used in personal modeling. Since this paper focuses on human preferences for embodied agents and there remains a lack of systematic analysis on which profiles correlate with such preferences, we consider human profiles across six dimensions potentially related to human expectations of embodied agents, spanning \emph{basic information}, \emph{robot experience}, \emph{robot trust}, \emph{personality traits}, \emph{autonomy}, and \emph{empathy} \citep{davis1983empathy,weinstein2012autonomy}. We then investigate whether these characteristics can assist in addressing Personalized SPI.

}

\begin{figure}[t]
    \centering
    \includegraphics[width=\linewidth]{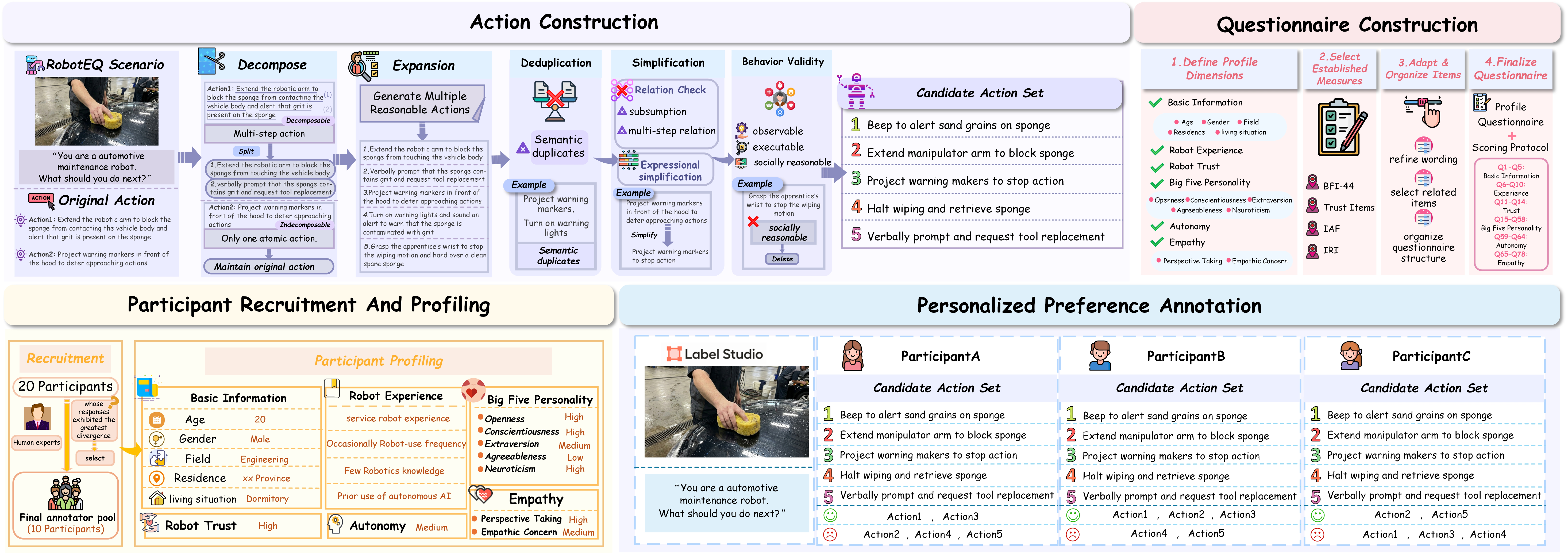}
    \caption{
    \lz{\textbf{Dataset construction}. \emph{1) Action construction}; \emph{2) Questionnaire construction}, to design a structured questionnaire that is related to human preferences for embodied agents; \emph{3) Participant recruitment and profiling}, to collect participant profiles via the above questionnaire and select diverse annotators; \emph{4) Personalized annotation}, to record each participant's preferred actions per scenario.}}
    \label{fig:dataset_overview}
\end{figure}

\lz{

\section{Dataset Construction}
\label{sec:Dataset_Construction}
Figure~\ref{fig:dataset_overview} summarizes our dataset construction pipeline. Building upon the scenarios and actions from prior works \citep{roboteq2026}, we first generate multiple socially appropriate candidate actions for each scenario. We then design a structured questionnaire to characterize participant traits related to their preferences for embodied agents. Next, we collect participant profiles and record each participant's preferred actions per scenario, yielding a dataset that establishes explicit links between human profiles and decision-making in SPI.

}

\lz{

\subsection{Action Construction}
\label{sec:action_construction}

This section describes our candidate action construction process. RobotEQ~\citep{roboteq2026} provides diverse scenarios with multiple proper actions per scenario. However, the original actions in RobotEQ vary in granularity, where some are concrete and detailed while others are brief. Such granularity differences may bias human preference judgments \citep{lian2026emoprefer, prefer1}. Therefore, we construct a unified action space where candidates are semantically distinct yet comparable in descriptive granularity. Specifically, we first decompose composite or multi-step actions into atomic units. We then expand each atomic action into at least five candidate behaviors to increase choice diversity. Next, we perform semantic deduplication to remove redundant or highly overlapping candidates. We further simplify actions by identifying potential subsumption or multi-step relations among the expanded candidates. Finally, we filter actions based on behavioral validity, requiring each candidate to be observable, executable, and socially reasonable. The resulting candidate space offers diverse, granularly balanced actions for personalized selection. For each scenario $s_i$, we denote its candidate action list as $\mathcal{A}_i \subseteq \mathcal{A}$, where $\mathcal{A}$ is the full set of valid actions. Detailed prompts for action construction are provided in Appendix \ref{app:action_prompts}.

}

\begin{figure}[t]
\centering

\captionsetup[subfigure]{
    font=scriptsize,
    skip=3pt,
    justification=centering,
    singlelinecheck=true
}

\newcommand{\panelheight}{1.12in}

\begin{subfigure}[t]{0.20\linewidth}
\centering
\vspace{0pt}

\begin{minipage}[t][\panelheight][c]{\linewidth}
\centering

\begingroup
\color{black}
\arrayrulecolor{black}
\scriptsize
\renewcommand{\arraystretch}{1.35}
\setlength{\tabcolsep}{1.1pt}

\resizebox{\linewidth}{!}{%
\begin{tabular}{@{}l@{\hspace{0.35em}}r@{}}
\specialrule{0.8pt}{0pt}{0.8pt}
\specialrule{0.3pt}{0pt}{2.2pt}
\textbf{Statistic} & \textbf{Value} \\
\midrule
\# of participants & 10 \\
\# of scenarios & 1,800 \\
Total annotations & 18,000 \\
\midrule
Action candidates per scenario & 5 \\
Mean selected actions per scenario & 2.24 \\
\specialrule{0.3pt}{2.2pt}{0.8pt}
\specialrule{0.8pt}{0pt}{0pt}
\end{tabular}
}

\endgroup
\end{minipage}

\caption{Main statistics}
\label{fig:dataset_statistics_table}
\end{subfigure}
\hfill
%
\begin{subfigure}[t]{0.20\linewidth}
\centering
\vspace{0pt}

\begin{minipage}[t][\panelheight][c]{\linewidth}
\centering

\includegraphics[
    width=0.88\linewidth,
    keepaspectratio
]{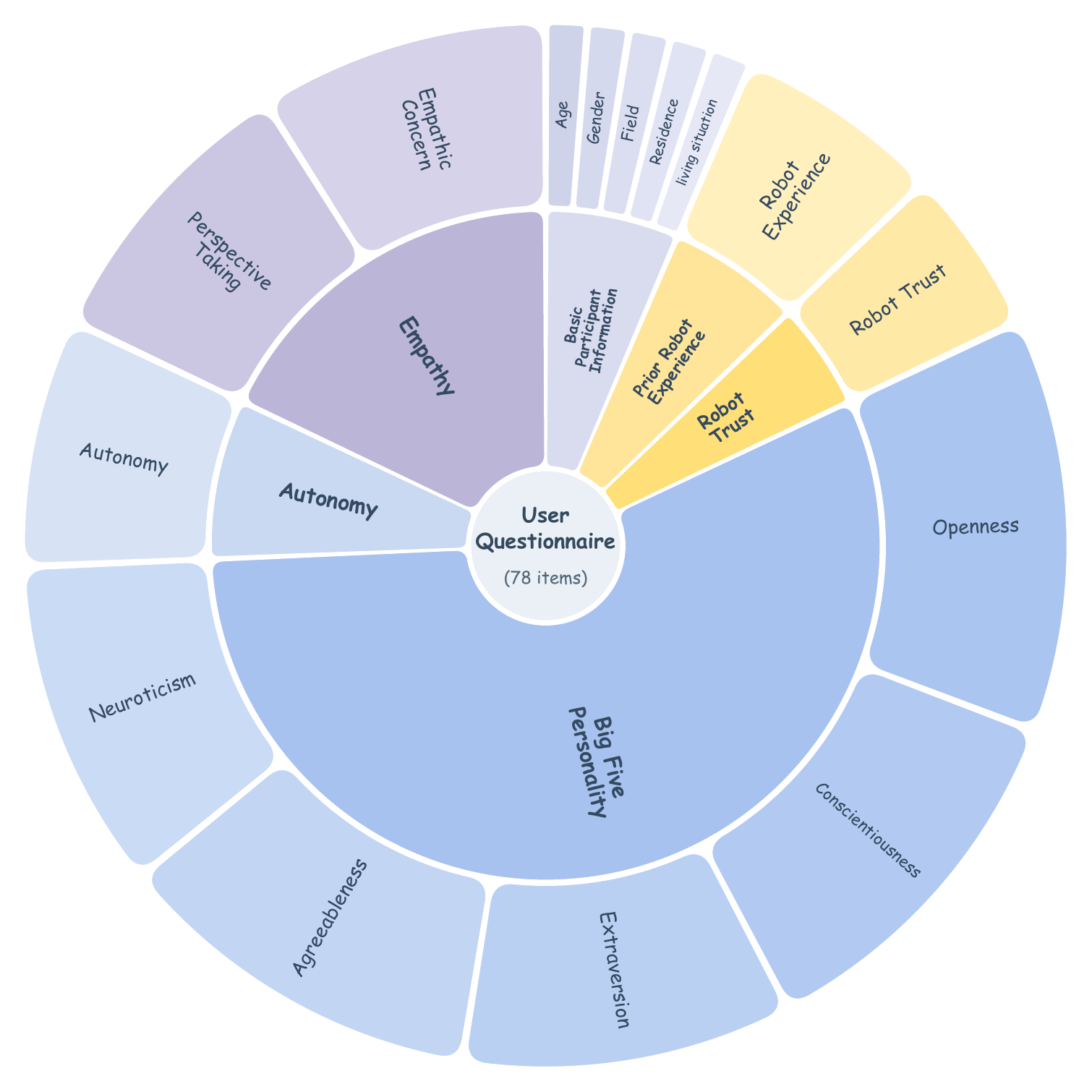}

\end{minipage}

\caption{Questionnaire}
\label{fig:questionnaire_structure}
\end{subfigure}
\hfill
%
\begin{subfigure}[t]{0.30\linewidth}
\centering
\vspace{0pt}

\begin{minipage}[t][\panelheight][c]{\linewidth}
\centering

\includegraphics[
    width=\linewidth,
    trim={0.46in 0 0.20in 0.56in},
    clip,
    keepaspectratio
]{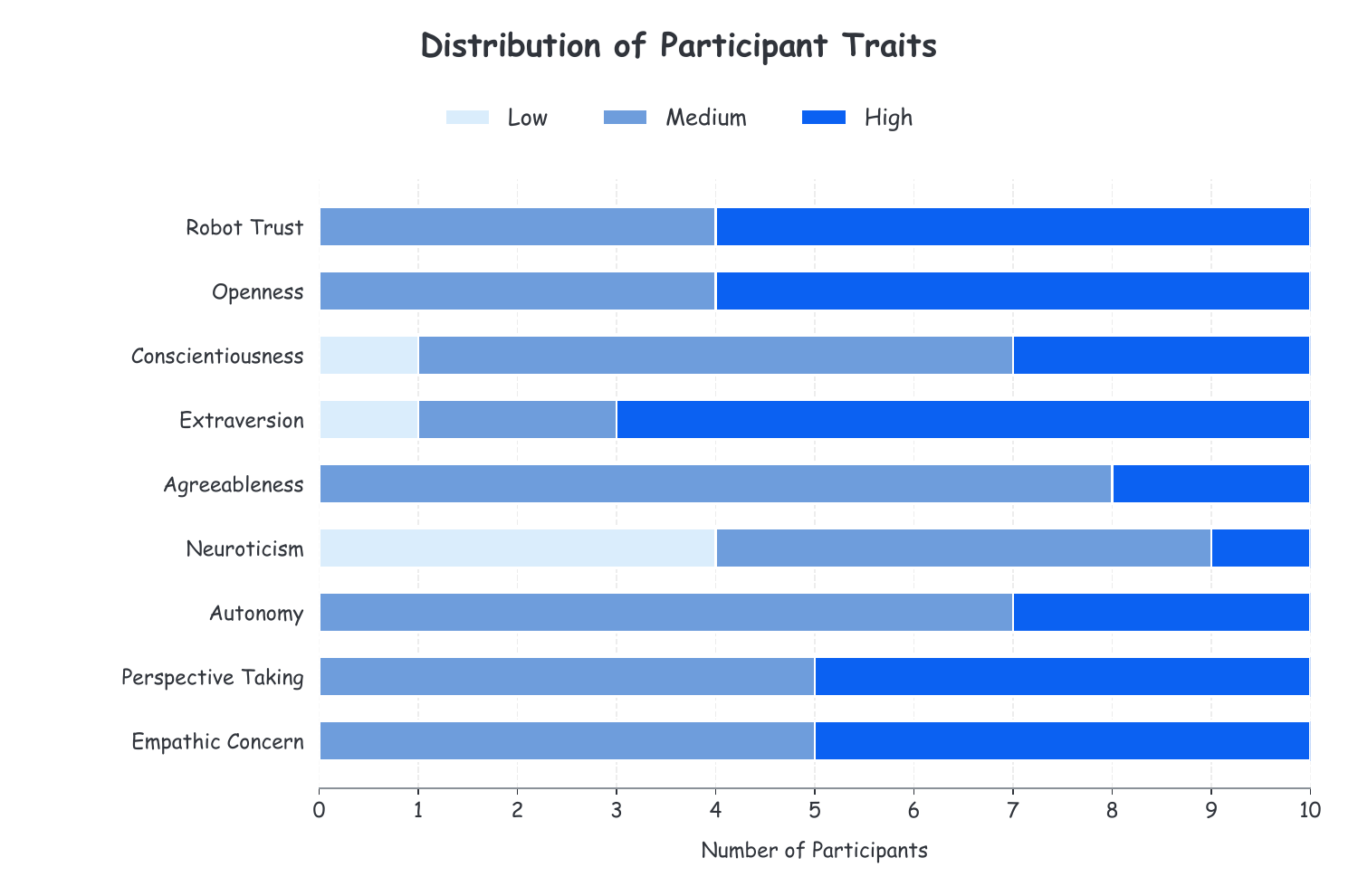}

\end{minipage}

\caption{Trait distribution}
\label{fig:participant_trait_distribution}
\end{subfigure}
\hfill
%
\begin{subfigure}[t]{0.26\linewidth}
\centering
\vspace{0pt}

\begin{minipage}[t][\panelheight][c]{\linewidth}
\centering

\includegraphics[
    width=\linewidth,
    trim={0 0 0.34in 0},
    clip,
    keepaspectratio
]{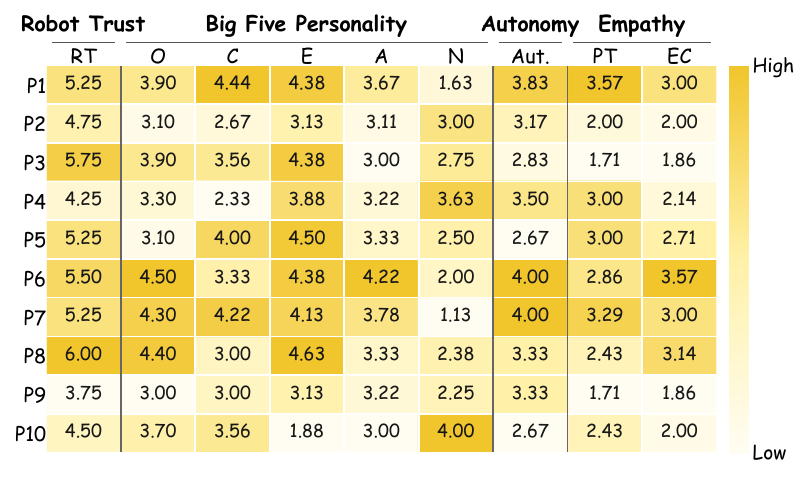}

\end{minipage}

\caption{Profile heatmap}
\label{fig:participant_profile_heatmap}
\end{subfigure}

\caption{
\lz{\textbf{Dataset statistics.}
(a) Overview of main statistics.
(b) Structure of the 78-item questionnaire.
(c) Distribution of participant traits.
(d) Heatmap of participant profiles.}
}
\label{fig:dataset_statistics}

\end{figure}

\arrayrulecolor{black}

\lz{

\subsection{Questionnaire Construction}
\label{sec:questionnaire_construction}

To measure participant traits associated with their preferences for embodied agents, we construct a structured questionnaire organized into six dimensions: \emph{basic information}, \emph{personality}, \emph{robot experience}, \emph{robot trust}, \emph{autonomy}, and \emph{empathy}. For personality, we adopt the widely used Big Five model~\citep{john1991bfi}, comprising Openness, Conscientiousness, Extraversion, Agreeableness, and Neuroticism. For robot trust, we adapt items from established trust measures~\citep{jian2000trust}. For autonomy, we administer the Self-Congruence and Low Susceptibility to Control subscales of the Index of Autonomous Functioning (IAF)~\citep{weinstein2012autonomy}. For empathy, we use the Perspective Taking and Empathic Concern subscales of the Interpersonal Reactivity Index (IRI)~\citep{davis1983empathy}. Full questionnaire items, response scales, and scoring rules are provided in Appendix \ref{app:questionnaire-scoring}.

}




\lz{

\subsection{Participant Recruitment and Profiling}
\label{sec:user_profiling}

We recruited twenty participants and asked them to complete the questionnaire. Human experts then manually selected the ten participants whose responses exhibited the greatest divergence, forming the final annotator pool. Figure \ref{fig:dataset_overview} shows a participant profile. Basic information includes age, gender, field of study, long-term residence, and living situation. Robot experience captures prior exposure to robotic systems, usage frequency, robotics knowledge, robot acceptance, and prior interaction with autonomous AI systems. Robot trust measures perceived reliability, trustworthiness, and confidence in such systems. The remaining components span the five personality traits, two autonomy dimensions, and two empathy dimensions. We represent the profile of participant $u$ as an ordered tuple $\mathbf{P}_u=(\mathbf{D}_u,\mathbf{R}_u,\mathbf{T}_u,\mathbf{B}_u,\mathbf{A}_u,\mathbf{E}_u)$, where $\mathbf{D}_u$ = basic information, $\mathbf{R}_u$ = robot experience, $\mathbf{T}_u$ = robot trust, $\mathbf{B}_u$ = Big Five personality, $\mathbf{A}_u$ = autonomy, and $\mathbf{E}_u$ = empathy.

}




\lz{

\subsection{Personalized Annotation}
\label{sec:preference_annotation}

To capture individual nuances, we present all participants with the same scenario and candidate action set. For each scenario, the annotation interface displays the scene image, a question, and five candidate actions. Details of the interface are provided in Appendix \ref{app:annotation-interface}. During annotation, participants select one action when they have a clear preference and select multiple actions when several candidates are equally preferred. Thus, each annotator's selection $Y_{u,i} \subseteq \mathcal{A}_i$ for participant $u$ and scenario $s_i$ is represented as a set rather than a single value. Each annotation is paired with the participant's profile $\mathbf{P}_u$, yielding participant-specific labels for identical scenarios and candidate sets. To assess participant reliability, a subset of scenarios is repeated during annotation without being marked as repeated items. The consistency of repeated scenarios reflects participant reliability.

}

\lz{

\subsection{Dataset Statistics}
\label{sec:dataset_statistics}

Figure~\ref{fig:dataset_statistics_table} presents the main statistics of our dataset, comprising 1,800 scenarios annotated by 10 participants, yielding 18K annotations. Each scenario contains 5 action candidates, with participants selecting an average of 2.24 actions per scenario. Beyond these statistics, the questionnaire structure in Figure~\ref{fig:questionnaire_structure} reveals how we profile each participant, covering basic information, robot experience, robot trust, personality, autonomy, and empathy. The trait distribution in Figure~\ref{fig:participant_trait_distribution} shows that our participants span a wide range of personal profiles. This diversity becomes most apparent in the heatmap in Figure~\ref{fig:participant_profile_heatmap}, where each participant displays a distinct combination of traits. These individual differences form the foundation of our personalized benchmark.

}

\cxy{\section{Experimental Setup}
\label{sec:experimental_setup}

}

\lz{
\subsection{Evaluation Metrics}
\label{sec:evaluation_metrics}
This section describes our evaluation metrics. Let $Y_{u,i}$ denote the ground-truth actions selected by participant $u$ for scenario $i$, $\hat{Y}_{u,i}$ the corresponding model prediction, and $N$ the total number of scenarios. For each participant $u$, we evaluate performance using two metrics. Set-level F-score $\mathrm{F}_{u}$ measures full-set matching, and $\mathrm{HIT}_{u}$ measures whether the model identifies at least one valid choice. To integrate these metrics, we propose the Preference Alignment Score (PAS), defined as:
\begin{equation}
\mathrm{F}_{u} = \frac{1}{N}\sum_{i=1}^N\frac{2|\hat{Y}_{u,i}\cap Y_{u,i}|}{|\hat{Y}_{u,i}|+|Y_{u,i}|},
\;
\mathrm{HIT}_{u} = \frac{1}{N}\sum_{i=1}^N\mathbf{1}(Y_{u,i}\cap\hat{Y}_{u,i}\neq\emptyset),
\;
\mathrm{PAS}_{u} = \frac{1}{2}F_{u} + \frac{1}{2}\mathrm{HIT}_{u}.
\end{equation}
The overall metric is computed as a weighted average across participants, with weights $w_u$ reflecting each participant's annotation reliability. Specifically, each participant annotates 20 samples twice, and $w_u$ is derived from the intra-participant agreement across these repeated items (see Appendix~\ref{app:annotation_consistency}).
\begin{equation}
\mathrm{M}_{\mathrm{overall}}
=
\frac{\sum_{u} w_u \mathrm{M}_u}
{\sum_{u} w_u},
\;
\mathrm{M}\in\{\mathrm{F},\mathrm{HIT},\mathrm{PAS}\}.
\label{eq:overall_metric}
\end{equation}

}









\begin{table*}[t!]
\centering



\caption{\lz{\textbf{Profile-conditioned evaluation.} This table presents the impact of incorporating human profiles on personalized SPI. \emph{No Profile} denotes results obtained without any profile information, while \emph{Best Profile} reports results under the optimal profile setting for each participant-model pair. Individual PAS is reported, and the Overall columns report F-score, HIT, and PAS. These metrics are defined in Section \ref{sec:evaluation_metrics}. $\Delta$ denotes the PAS improvement achieved with profile conditioning.}}

\label{tab:main_results}

\vspace{3pt}

\fontsize{7.2}{8.0}\selectfont
\renewcommand{\arraystretch}{0.94}
\setlength{\tabcolsep}{1.65pt}

\resizebox{\textwidth}{!}{%
\begin{tabular}{
@{}
l
l
cccccccccc
cccc
@{}
}

\specialrule{0.95pt}{0pt}{1.1pt}
\specialrule{0.35pt}{0pt}{4.0pt}

\multirow{2}{*}{\textbf{Model}}
&
\multirow{2}{*}{\textbf{Setting}}
&
\multicolumn{10}{c}{\textbf{Individual PAS (\%)}}
&
\multicolumn{4}{c}{\textbf{Overall}}
\\[-1pt]

\cmidrule(lr){3-12}
\cmidrule(lr){13-16}

&
&
\textbf{P$_1$}
&
\textbf{P$_2$}
&
\textbf{P$_3$}
&
\textbf{P$_4$}
&
\textbf{P$_5$}
&
\textbf{P$_6$}
&
\textbf{P$_7$}
&
\textbf{P$_8$}
&
\textbf{P$_9$}
&
\textbf{P$_{10}$}
&
\textbf{F}
&
\textbf{HIT}
&
\textbf{PAS}
&
\textbf{$\Delta$}
\\[1pt]

\specialrule{0.35pt}{2pt}{1pt}
\specialrule{0.35pt}{0pt}{2.0pt}

\multirow{2}{*}{Qwen3.5-4B-Instruct} & No Profile
& 58.77 & 56.57 & 46.98 & 54.27 & 49.42 & 59.17 & 48.01 & 63.24 & 50.72 & 57.20
& 41.79 & 67.32 & \textbf{54.56} & \\

\arrayrulecolor{gray!55}\cmidrule(lr){2-15}\arrayrulecolor{black}

\rowcolor{yellow!18}\cellcolor{white} & \textbf{Best Profile}
& 78.42 & 71.81 & 56.86 & 67.04 & 63.03 & 78.54 & 73.60 & 84.66 & 58.52 & 77.46
& 57.14 & 85.60 & \textbf{71.37} & +16.81 \\

\specialrule{0.30pt}{1.3pt}{1.3pt}

\multirow{2}{*}{Qwen3-VL-30B-A3B-Instruct} & No Profile
& 54.19 & 52.86 & 45.93 & 50.51 & 47.48 & 55.42 & 42.57 & 57.91 & 51.06 & 53.12
& 38.97 & 63.37 & \textbf{51.17} & \\

\arrayrulecolor{gray!55}\cmidrule(lr){2-15}\arrayrulecolor{black}

\rowcolor{yellow!18}\cellcolor{white} & \textbf{Best Profile}
& 76.53 & 66.26 & 47.06 & 64.38 & 59.25 & 77.56 & 70.19 & 80.73 & 57.04 & 76.23
& 55.14 & 80.57 & \textbf{67.86} & +16.69 \\

\specialrule{0.30pt}{1.3pt}{1.3pt}

\multirow{2}{*}{InternVL3.5-30B-A3B-Instruct} & No Profile
& 50.81 & 55.62 & 47.92 & 50.34 & 47.98 & 54.17 & 45.31 & 57.61 & 50.95 & 54.64
& 39.45 & 63.78 & \textbf{51.62} & \\

\arrayrulecolor{gray!55}\cmidrule(lr){2-15}\arrayrulecolor{black}

\rowcolor{yellow!18}\cellcolor{white} & \textbf{Best Profile}
& 70.50 & 66.35 & 54.20 & 61.34 & 60.48 & 72.33 & 68.62 & 80.66 & 55.88 & 74.88
& 54.38 & 79.51 & \textbf{66.94} & +15.32 \\

\specialrule{0.30pt}{1.3pt}{1.3pt}

\multirow{2}{*}{Qwen3-VL-4B-Instruct} & No Profile
& 58.42 & 57.32 & 46.09 & 54.23 & 51.27 & 59.31 & 49.60 & 63.12 & 50.61 & 58.18
& 42.16 & 67.64 & \textbf{54.90} & \\

\arrayrulecolor{gray!55}\cmidrule(lr){2-15}\arrayrulecolor{black}

\rowcolor{yellow!18}\cellcolor{white} & \textbf{Best Profile}
& 73.74 & 70.45 & 49.30 & 62.48 & 62.16 & 75.09 & 69.21 & 80.85 & 54.90 & 74.27
& 53.83 & 81.18 & \textbf{67.51} & +12.60 \\

\specialrule{0.30pt}{1.3pt}{1.3pt}

\multirow{2}{*}{MiMo-Embodied-7B} & No Profile
& 64.05 & 61.06 & 53.12 & 58.95 & 53.94 & 65.06 & 54.03 & 65.93 & 56.02 & 62.78
& 45.68 & 73.52 & \textbf{59.60} & \\

\arrayrulecolor{gray!55}\cmidrule(lr){2-15}\arrayrulecolor{black}

\rowcolor{yellow!18}\cellcolor{white} & \textbf{Best Profile}
& 77.01 & 71.25 & 58.12 & 67.08 & 61.06 & 79.00 & 75.20 & 85.30 & 59.51 & 79.12
& 58.11 & 85.38 & \textbf{71.74} & +12.14 \\

\specialrule{0.30pt}{1.3pt}{1.3pt}

\multirow{2}{*}{MiMo-VL-7B-SFT-2508} & No Profile
& 65.03 & 63.53 & 55.12 & 60.73 & 55.79 & 66.46 & 57.82 & 69.89 & 57.08 & 65.39
& 47.70 & 76.03 & \textbf{61.86} & \\

\arrayrulecolor{gray!55}\cmidrule(lr){2-15}\arrayrulecolor{black}

\rowcolor{yellow!18}\cellcolor{white} & \textbf{Best Profile}
& 76.98 & 75.85 & 63.46 & 68.63 & 62.77 & 77.62 & 78.05 & 86.32 & 58.98 & 80.22
& 60.00 & 86.74 & \textbf{73.37} & +11.51 \\

\specialrule{0.30pt}{1.3pt}{1.3pt}

\multirow{2}{*}{Qwen3.5-9B-Instruct} & No Profile
& 61.43 & 57.94 & 43.98 & 56.35 & 51.52 & 62.53 & 49.92 & 64.65 & 50.27 & 59.27
& 42.48 & 69.21 & \textbf{55.84} & \\

\arrayrulecolor{gray!55}\cmidrule(lr){2-15}\arrayrulecolor{black}

\rowcolor{yellow!18}\cellcolor{white} & \textbf{Best Profile}
& 74.11 & 69.09 & 49.41 & 63.62 & 60.19 & 73.91 & 69.06 & 80.43 & 55.96 & 73.79
& 53.70 & 80.81 & \textbf{67.25} & +11.41 \\

\specialrule{0.30pt}{1.3pt}{1.3pt}

\multirow{2}{*}{Qwen3.5-27B-Instruct} & No Profile
& 69.57 & 63.97 & 57.44 & 63.73 & 56.11 & 69.73 & 56.49 & 71.79 & 62.47 & 68.15
& 50.45 & 77.81 & \textbf{64.13} & \\

\arrayrulecolor{gray!55}\cmidrule(lr){2-15}\arrayrulecolor{black}

\rowcolor{yellow!18}\cellcolor{white} & \textbf{Best Profile}
& 80.38 & 72.89 & 59.81 & 69.30 & 63.32 & 79.70 & 76.56 & 83.71 & 66.39 & 82.46
& 61.12 & 86.58 & \textbf{73.85} & +9.72 \\

\specialrule{0.30pt}{1.3pt}{1.3pt}

\multirow{2}{*}{MiMo-VL-7B-RL-2508} & No Profile
& 68.75 & 66.75 & 59.07 & 63.96 & 58.25 & 69.79 & 63.04 & 73.45 & 61.88 & 68.58
& 50.38 & 80.79 & \textbf{65.58} & \\

\arrayrulecolor{gray!55}\cmidrule(lr){2-15}\arrayrulecolor{black}

\rowcolor{yellow!18}\cellcolor{white} & \textbf{Best Profile}
& 76.24 & 75.99 & 67.63 & 71.59 & 64.62 & 78.19 & 77.76 & 85.97 & 64.75 & 80.95
& 59.31 & 90.33 & \textbf{74.82} & +9.24 \\

\specialrule{0.30pt}{1.3pt}{1.3pt}

\multirow{2}{*}{VideoChat2-4B} & No Profile
& 56.79 & 58.01 & 43.81 & 53.65 & 48.69 & 56.58 & 54.49 & 67.07 & 42.23 & 58.77
& 40.96 & 67.59 & \textbf{54.28} & \\

\arrayrulecolor{gray!55}\cmidrule(lr){2-15}\arrayrulecolor{black}

\rowcolor{yellow!18}\cellcolor{white} & \textbf{Best Profile}
& 64.87 & 67.07 & 54.65 & 59.38 & 56.95 & 67.33 & 64.57 & 75.42 & 51.60 & 69.59
& 50.13 & 76.83 & \textbf{63.48} & +9.20 \\

\specialrule{0.30pt}{1.3pt}{1.3pt}

\multirow{2}{*}{GLM-4.6V-Flash} & No Profile
& 55.04 & 57.39 & 49.73 & 52.08 & 49.97 & 55.68 & 49.38 & 62.79 & 52.00 & 58.20
& 41.97 & 66.85 & \textbf{54.41} & \\

\arrayrulecolor{gray!55}\cmidrule(lr){2-15}\arrayrulecolor{black}

\rowcolor{yellow!18}\cellcolor{white} & \textbf{Best Profile}
& 64.51 & 65.23 & 49.19 & 57.04 & 56.88 & 65.11 & 65.79 & 74.41 & 51.80 & 69.01
& 49.17 & 75.29 & \textbf{62.23} & +7.82 \\

\specialrule{0.30pt}{1.3pt}{1.3pt}

\multirow{2}{*}{InternVL3.5-8B-Instruct} & No Profile
& 62.67 & 60.50 & 55.28 & 56.62 & 52.51 & 62.03 & 53.72 & 67.18 & 54.01 & 63.80
& 45.85 & 72.28 & \textbf{59.06} & \\

\arrayrulecolor{gray!55}\cmidrule(lr){2-15}\arrayrulecolor{black}

\rowcolor{yellow!18}\cellcolor{white} & \textbf{Best Profile}
& 71.76 & 65.39 & 53.52 & 59.03 & 57.39 & 70.74 & 68.73 & 78.34 & 54.12 & 71.54
& 52.25 & 78.73 & \textbf{65.49} & +6.43 \\

\specialrule{0.30pt}{1.3pt}{1.3pt}

\multirow{2}{*}{Gemma-3-12B} & No Profile
& 65.13 & 61.41 & 47.46 & 60.16 & 55.44 & 65.19 & 56.01 & 69.75 & 51.82 & 64.68
& 45.37 & 74.28 & \textbf{59.83} & \\

\arrayrulecolor{gray!55}\cmidrule(lr){2-15}\arrayrulecolor{black}

\rowcolor{yellow!18}\cellcolor{white} & \textbf{Best Profile}
& 70.72 & 64.72 & 53.23 & 63.74 & 59.51 & 70.53 & 67.58 & 76.81 & 54.88 & 72.40
& 50.96 & 80.48 & \textbf{65.72} & +5.90 \\

\specialrule{0.30pt}{1.3pt}{1.3pt}

\multirow{2}{*}{Kimi-VL-A3B-Instruct} & No Profile
& 54.24 & 56.26 & 42.33 & 52.89 & 51.08 & 57.25 & 49.44 & 62.50 & 46.64 & 56.04
& 40.07 & 65.80 & \textbf{52.94} & \\

\arrayrulecolor{gray!55}\cmidrule(lr){2-15}\arrayrulecolor{black}

\rowcolor{yellow!18}\cellcolor{white} & \textbf{Best Profile}
& 58.52 & 61.03 & 44.02 & 55.30 & 54.88 & 62.02 & 62.57 & 69.34 & 48.22 & 64.47
& 45.26 & 71.30 & \textbf{58.28} & +5.34 \\

\specialrule{0.30pt}{1.3pt}{1.3pt}

\multirow{2}{*}{VideoLLaMA3-7B} & No Profile
& 51.89 & 52.32 & 35.62 & 49.23 & 47.77 & 56.16 & 41.49 & 55.44 & 41.49 & 50.76
& 35.86 & 60.35 & \textbf{48.11} & \\

\arrayrulecolor{gray!55}\cmidrule(lr){2-15}\arrayrulecolor{black}

\rowcolor{yellow!18}\cellcolor{white} & \textbf{Best Profile}
& 57.33 & 56.77 & 35.03 & 50.97 & 50.95 & 60.98 & 51.86 & 64.40 & 42.20 & 57.96
& 39.53 & 66.31 & \textbf{52.92} & +4.81 \\

\specialrule{0.30pt}{1.3pt}{1.3pt}

\multirow{2}{*}{Qwen3-VL-8B-Instruct} & No Profile
& 67.62 & 63.06 & 55.51 & 60.70 & 55.37 & 67.61 & 56.38 & 70.24 & 56.90 & 65.66
& 47.96 & 76.21 & \textbf{62.08} & \\

\arrayrulecolor{gray!55}\cmidrule(lr){2-15}\arrayrulecolor{black}

\rowcolor{yellow!18}\cellcolor{white} & \textbf{Best Profile}
& 72.68 & 66.14 & 51.13 & 60.82 & 58.71 & 71.32 & 67.86 & 75.76 & 56.39 & 70.56
& 50.95 & 79.89 & \textbf{65.42} & +3.34 \\

\specialrule{0.30pt}{1.3pt}{1.3pt}

\multirow{2}{*}{Qwen2.5-VL} & No Profile
& 63.83 & 62.92 & 52.44 & 60.80 & 56.87 & 65.29 & 56.71 & 71.44 & 55.46 & 65.81
& 47.17 & 75.50 & \textbf{61.33} & \\

\arrayrulecolor{gray!55}\cmidrule(lr){2-15}\arrayrulecolor{black}

\rowcolor{yellow!18}\cellcolor{white} & \textbf{Best Profile}
& 66.50 & 63.66 & 53.52 & 60.60 & 58.11 & 68.80 & 66.43 & 73.92 & 53.28 & 68.63
& 49.16 & 78.14 & \textbf{63.65} & +2.32 \\

\specialrule{0.30pt}{1.3pt}{1.3pt}

\multirow{2}{*}{InternVL3.5-4B-Instruct} & No Profile
& 61.31 & 62.70 & 55.97 & 57.51 & 54.64 & 63.20 & 56.78 & 68.66 & 55.28 & 63.67
& 46.88 & 73.52 & \textbf{60.20} & \\

\arrayrulecolor{gray!55}\cmidrule(lr){2-15}\arrayrulecolor{black}

\rowcolor{yellow!18}\cellcolor{white} & \textbf{Best Profile}
& 66.12 & 61.72 & 55.38 & 56.49 & 55.13 & 67.23 & 64.48 & 75.51 & 52.69 & 65.17
& 49.08 & 75.81 & \textbf{62.44} & +2.25 \\

\specialrule{0.35pt}{3pt}{1pt}
\specialrule{0.95pt}{0pt}{0pt}

\end{tabular}%
}

\end{table*}


\lz{

\subsection{Benchmarking Candidates}
Our benchmark evaluates a range of representative vision-language models (VLMs). All models support vision-language reasoning, which mimics how VLMs think as embodied agents. Detailed prompts are provided in Appendix \ref{app:benchmarking-prompts}. For each model, we inject distinct personality profiles as conditional inputs and evaluate the model's responses under each configuration. Due to the high cost of our large-scale personalized evaluation, we focus on open-source models with state-of-the-art multimodal capabilities. Exploration of closed-source models is left to our future work.

}

\lz{

\section{Results and Discussion}
This is the first benchmark designed for personalized SPI. In this section, we first reveal the role of profiles in preference prediction. Then, we conduct an ablation study, revealing the impact of scenario images and few-shot prompting. Next, we investigate the relationship between trait similarity and preference consistency and conduct a trait-level analysis. Meanwhile, we introduce a self-evolving-driven solution to automatically capture profile-aware experience for addressing SPI. Finally, we conduct a case study for visualizing the benefits of participant profiles.

\subsection{Profile-Conditioned Evaluation}
This section examines whether individual traits can improve preference prediction. Results are presented in Table~\ref{tab:main_results}. Under identical scenarios, questions, and candidate actions, we compare two input settings. \textbf{1) No Profile}, where the model receives no participant-specific profile and serves as the baseline; \textbf{2) Best Profile}, where human profiles are provided as additional inputs, and we report the optimal profile setting for each participant-model pair. More results on the impact of individual profiles are provided in Appendix~\ref{app:participant_profile_results}. As shown in Table~\ref{tab:main_results}, profile-conditioned models improve overall PAS scores across all models, though the magnitude of improvement varies. Qwen3.5-4B-Instruct presents the largest relative gains, while Qwen3.5-27B-Instruct achieves the highest absolute profile-conditioned performance. These results confirm that participant profiles provide valuable signals for personalized preference prediction. The underlying reason is intuitive. Without profile information, a model must produce a single consensus answer averaged across all users, forcing it to ignore individual differences and default to the most generic option. Incorporating personality profiles supplies concrete cues about a specific user, enabling the model to distinguish among multiple reasonable actions and select the one aligned with that user's tendencies. Notably, we observe substantial variation in improvement across participants. Some show massive gains under profile conditioning, while others see only modest improvements. This suggests that the role of human traits varies for different participants.

}


\begin{figure*}[t]
    \centering
    \captionsetup[subfigure]{font=footnotesize,skip=2pt}
    \begin{subfigure}[t]{0.45\textwidth}
        \centering
        \includegraphics[width=\linewidth]{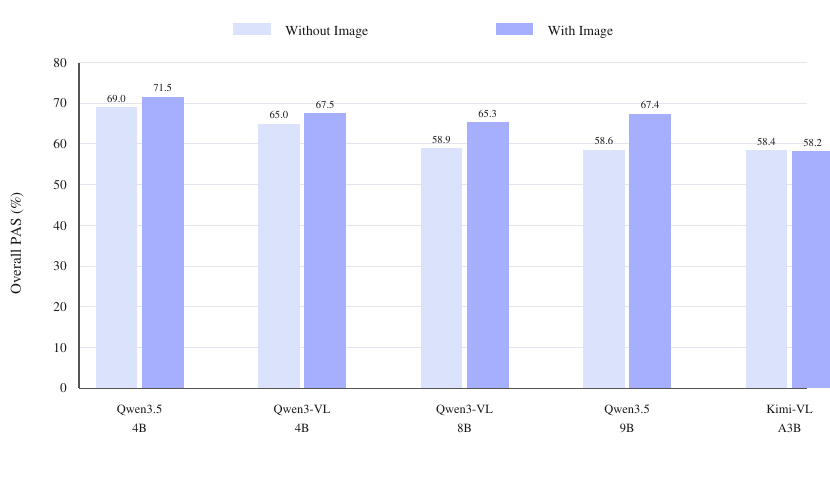}
        \caption{\lz{Impact of scenario images}}
        \label{fig:image_ablation}
    \end{subfigure}
    \hspace{0.06\textwidth}
    \begin{subfigure}[t]{0.45\textwidth}
        \centering
        \includegraphics[width=\linewidth]{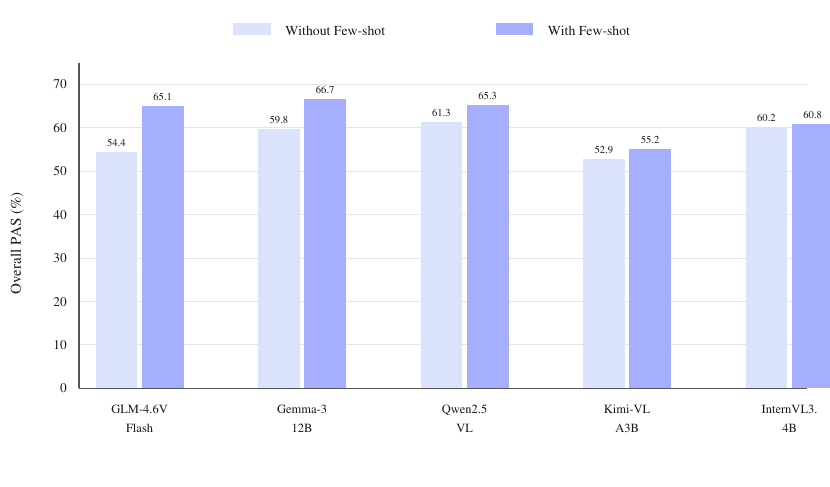}
        \caption{\lz{Impact of few-shot prompting}}
        \label{fig:fewshot_results}
    \end{subfigure}
    \caption{\lz{\textbf{Ablation study on different inputs.} (a) Compares performance with and without scenario images. (b) Compares performance with and without few-shot prompting.}}
    \label{fig:ablation_fewshot}
\end{figure*}

\lz{
\subsection{Ablation Study on Different Inputs}
\label{sec:context_demo_ablation}

\paragraph{Impact of Scenario Images.}
Figure~\ref{fig:image_ablation} evaluates whether visual scene information remains beneficial when participant profiles are available. We compare each model under its optimal profile setting. Experimental results confirm that visual cues substantially aid preference judgment. This directly validates the multimodal dependency defined in SPI~\citep{roboteq2026}, demonstrating that personalized preference prediction relies on both user-specific profiles and scenario context.

\paragraph{Impact of Few-shot Prompting.}
We further examine whether latent user preferences can be extracted from in-context demonstrations. Figure~\ref{fig:fewshot_results} evaluates the effect of few-shot prompting. Specifically, we extract 20 question–answer pairs from each participant's annotations as demonstration examples. Paired bars in Figure~\ref{fig:fewshot_results} compare each model with and without these examples, assessing whether such demonstrations carry useful signals. Experimental results demonstrate that few-shot prompting yields performance gains, confirming that beyond explicit personality profiles, in-context demonstrations provide complementary signals for personalized SPI.

}

\lz{
\subsection{Relationship between Trait Similarity and Preference Consistency}
\label{sec:profile_similarity_action_consistency}

Prior experiments confirm that incorporating human traits improves model performance on individual preference prediction. Here, we further investigate the relationship between trait similarity and preference consistency, with results presented in Figure~\ref{fig:profile_action_scatter_regression}. For each selected profile dimension, every panel visualizes all 45 participant pairs ($C_{10}^2=45$), with the x-axis indicating trait similarity and the y-axis indicating preference consistency. Trait similarity is quantified as the negative normalized difference between profile scores; preference consistency is measured by the F-score between two participants (Details of the caculations are shown in Appendix~\ref{app:annotation_consistency}). Experimental results demonstrate that \emph{Robot Trust} and \emph{Openness} exhibit positive correlations, whereas \emph{Conscientiousness} and \emph{Extraversion} show negative associations. These mixed directions indicate that profile similarity and preference consistency are not universally positively or negatively correlated. Therefore, different profile dimensions should be leveraged adaptively.

\begin{figure}[h]
    \centering
    \includegraphics[width=\textwidth]{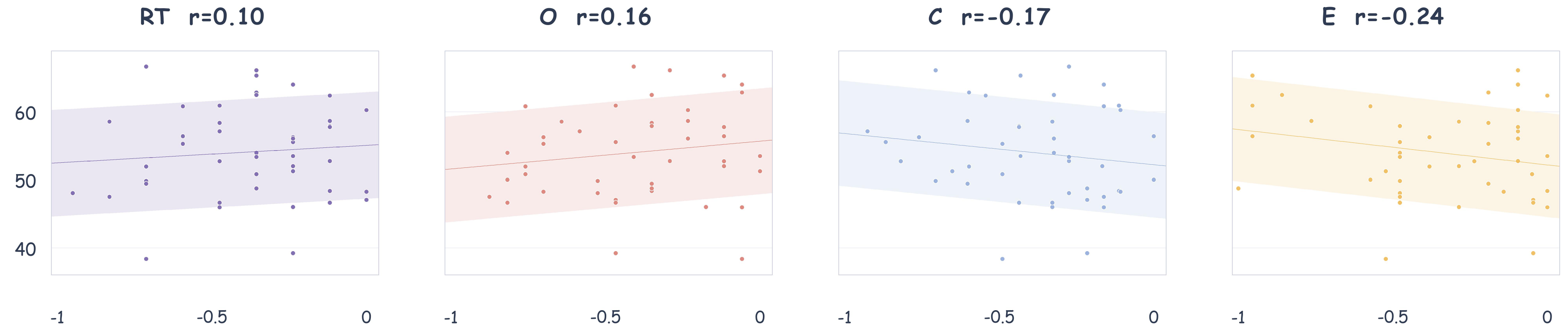}
    \caption{\lz{\textbf{Relationship between trait similarity and preference consistency.} In each panel, the x-axis indicates trait similarity and the y-axis indicates preference consistency. Each panel visualizes all 45 participant pairs ($C_{10}^2=45$) for a single profile dimension.}}
    \label{fig:profile_action_scatter_regression}
\end{figure}

}

\lz{

\subsection{Trait-Level Analysis}
\label{sec:trait_contribution}

\begin{figure}[!t]
    \centering
    \includegraphics[width=\textwidth]{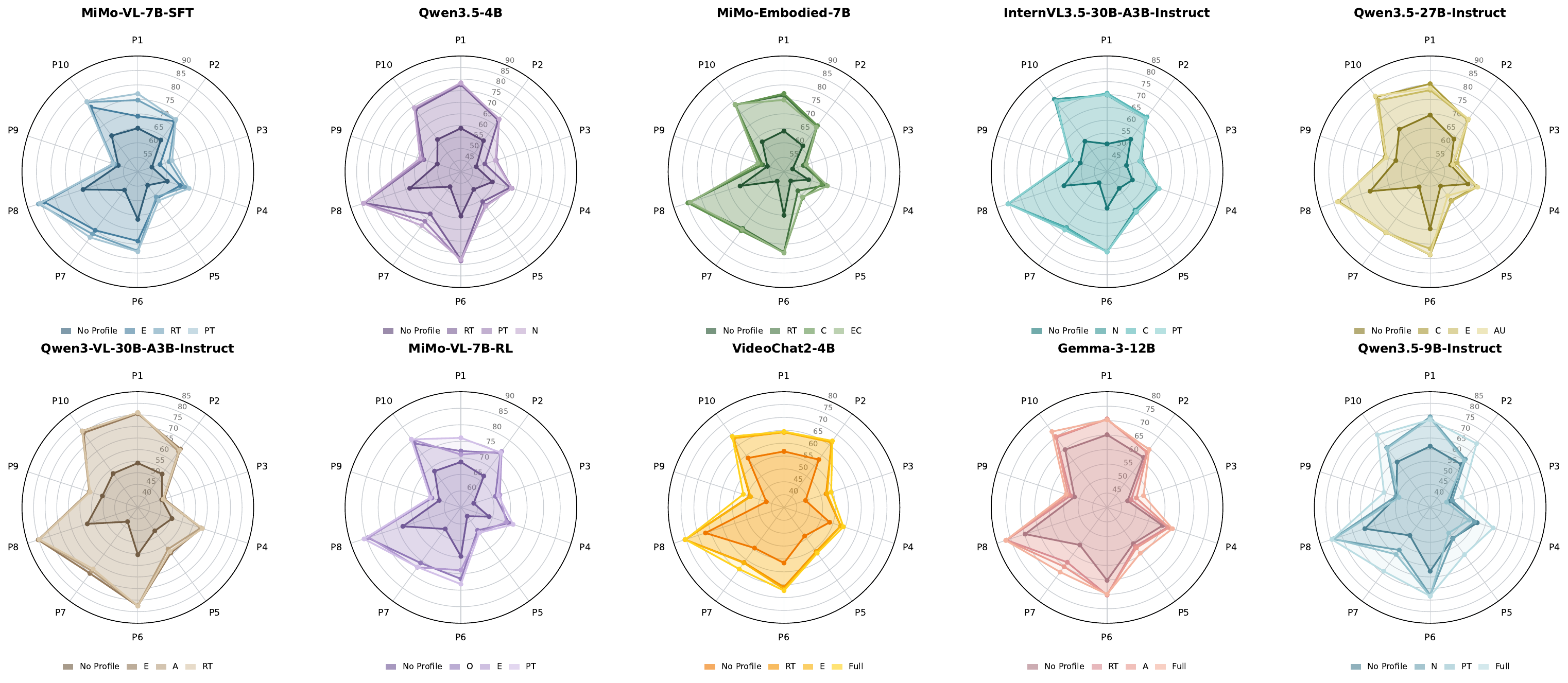}
    \caption{\lz{\textbf{Trait-level analysis.} Each panel shows results for one backbone model. Within each panel, we select the top-3 most informative traits and use the no-profile setting as the baseline, revealing the contribution of individual profile dimensions to personalized SPI.}}
    \label{fig:trait_radar}
\end{figure}

This section presents a trait-level analysis of how profile dimensions contribute to personalized prediction. Figure~\ref{fig:trait_radar} visualizes trait-level contributions across different backbone models. Results show that no single profile dimension consistently dominates. Some backbones benefit from robot-related or autonomy-related information, while others rely more on personality or empathy traits. Notably, the complete profile is not uniformly optimal. Incorporating all available user information does not guarantee improved performance and may even introduce noise. This indicates that profile selection is a model- and participant-dependent problem, rather than a simple case where more information is always better. Therefore, personalized SPI requires adaptation to the target model and user, rather than treating user profiles as a monolithic block.

}

\lz{

\subsection{Solution Exploration: Iterative Profile-Aware Experience Induction}
\label{sec:experience_refinement}

\begin{figure}[t]
    \centering
    \includegraphics[width=\textwidth]{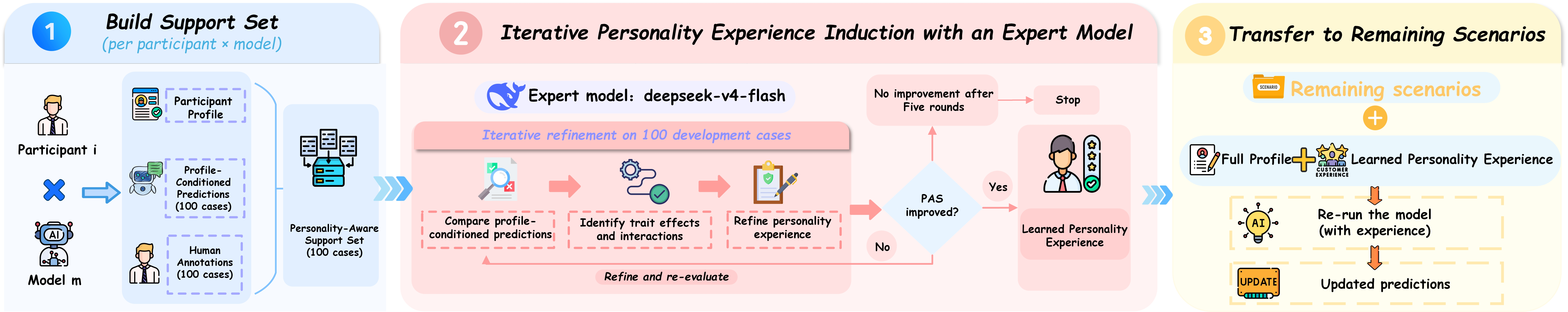}
    \caption{\lz{\textbf{Solution.} For each participant-model combination, we analyze prediction errors, iteratively induce profile-related experience, and transfer these experiences to the remaining scenarios.}}
    \label{fig:experience_transfer}
\end{figure}

\begin{wraptable}{R}{0.48\textwidth}
\centering
\resizebox{0.46\textwidth}{!}{
\small
\setlength{\tabcolsep}{5pt}
\renewcommand{\arraystretch}{1.05}
\begin{tabular}{lcc}
\toprule
\textbf{Model} &
\textbf{w/ Best Profile} &
\textbf{w/ Experience} \\
\midrule
GLM-4.6V-Flash & 59.18 & 64.91 \\
InternVL3.5-4B-Instruct & 61.92 & 63.30 \\
Kimi-VL-A3B-Instruct & 54.97 & 60.65 \\
Qwen3.5-4B & 71.05 & 72.44 \\
Qwen3.5-9B-Instruct & 63.76 & 72.69 \\
\bottomrule
\end{tabular}
}
\caption{\lz{Performance comparison between Best Profile and Experience-augmented settings.}}
\label{tab:experience_pas_2}
\end{wraptable}
Prior experiments show that although participant profiles provide useful signals, we need to manually select proper profiles for each participant-model combination. In this section, we explore whether this manual selection process can be automated. To this end, we propose \textbf{Iterative Profile-Aware Experience Induction}, building on the idea of \emph{Self-Evolving} mechanism \citep{selfenvolving}, where a model continuously reflects on its own outputs to extract actionable insights. As shown in Figure~\ref{fig:experience_transfer}, for each participant-model combination, we first collect support cases, each containing the scenario, the participant's full profile, the model's prediction, and the human-annotated preference. We then employ an expert model to review prediction failures, examining recurring error patterns and how different trait combinations influence outcomes. From these observations, the expert model distills profile-aware experience. We repeat this experience induction process iteratively until no further performance improvement on the support set. The resulting experience is then applied to the remaining scenarios. Implementation details are provided in Appendix~\ref{app:implementation_details}. To evaluate the induced experience, we randomly sample 850 scenarios disjoint from the 100 support cases. Table~\ref{tab:experience_pas_2} compares the Best Profile setting against the experience-augmented setting. Experimental results demonstrate that appending the induced experience consistently improves personalized preference prediction. The gain ranges from 1.38 to 8.94 percentage points across models, with Qwen3.5-9B-Instruct showing the largest improvement. This proves that our method successfully distills profile-related insights from failures and generalizes to unseen scenarios without any model retraining. These findings confirm that our approach offers a practical path toward personalized SPI.

}

\lz{

\subsection{Case Study}
\label{sec:case_study}
Figure~\ref{fig:case_study} presents a case study illustrating how participant profiles influence action selection. The scenario depicts a wheelchair user navigating a crowded tourist attraction.
Participant P10's ground-truth preferences prioritize safety protection, accessible information, and consent-based assistance. Under the no-profile setting, most models predict $A_5$, treating the scenario as a generic tour-guidance task. After conditioning on the full profile, model predictions shift substantially toward P10's preferred actions. This case demonstrates that participant profiles alter both action selection and scene interpretation. Without participant profiles, models default to generic tour-guidance behaviors. With these profiles, predictions align with accessibility, safety, and appropriate assistance, better matching the participant's preferred balance between proactive support and user autonomy.
\begin{figure*}[t]
    \centering
    \includegraphics[width=\textwidth]{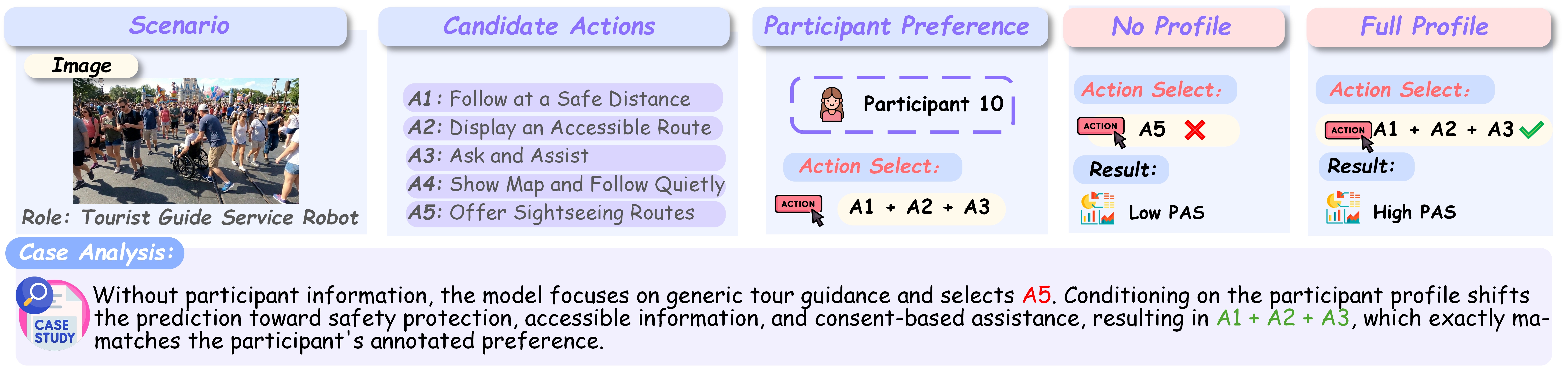}
    \caption{\lz{\textbf{Case study.} For the wheelchair-tourist scenario, participant P10 prefers actions emphasizing safety protection, accessible information, and consent-based assistance. Without profile information, the model selects the generic tour-guidance action $A_5$. conditioning on the full profile shifts predictions to $\{A_1,A_2,A_3\}$, aligning with P10's preferences.}}
    \label{fig:case_study}
\end{figure*}

}

\lz{

\section{Conclusion}
This paper extends SPI research from population-level modeling to individual-level analysis. To this end, we introduce \textbf{RobotEQ 3.0}, a benchmark that explicitly links human profiles to their preferred actions. Our benchmark comprises 1,800 scenarios, 10 participants with user profiles, and 18K individualized preference annotations. Experimental results demonstrate that human profiles provide valuable signals for personalized preference prediction. However, different profile dimensions contribute differently across participant-model combinations, and the complete profile is not uniformly optimal. These results suggest that personalized modeling must account for variation at the participant-model level. Therefore, we propose a self-evolution-driven method that extracts profile-related experience. This experience complements static profile information and improves personalized behavior prediction. This paper aims to shift SPI research beyond unified behavior selection for the average user to adapt to the preferences of specific individuals.

\paragraph{Limitations and Future Work.}
Due to funding constraints, our exploration of personalized SPI is limited to ten participants. The scale and diversity of the participant pool remain limited. In future work, we will cover a broader range of participants. Meanwhile, this paper characterizes individual differences through structured user profiles, which provide fine-grained descriptions of user characteristics. However, some relevant individual factors may still be missing. We plan to explore additional user dimensions linked to human preferences on embodied agents. Additionally, our current benchmark mainly includes participants from a Chinese cultural background. Future work will investigate cross-cultural effects on personalized SPI. Finally, due to the high computational cost of large-scale personalized evaluation, we focus on open-source models with state-of-the-art multimodal capabilities. Exploration of closed-source models is deferred to future work. 

}

\lz{

\section*{Ethics Statement}
This work involves human participants who completed questionnaires and provided annotations. All participants gave informed consent for the use of their annotation results. Annotators were compensated at approximately ¥150 per hour, a rate exceeding local standards. Our dataset is released under the CC BY-NC 4.0 license, which restricts use to non-commercial purposes and outlines guidelines for responsible use. Furthermore, the benchmark focuses exclusively on prosocial robot service scenarios, with all violent, discriminatory, or otherwise harmful content explicitly excluded. We emphasize that benchmark results do not constitute evidence of real-world deployment readiness. Further validation is required before any system is deployed in human environments.

}

\lz{

\section*{Reproducibility Statement}
The full dataset will be made publicly available upon paper acceptance. The dataset construction pipeline is detailed in Section \ref{sec:Dataset_Construction}. The appendix further provides prompt templates, representative cases, and annotation guidelines to facilitate replication. In summary, we have made every effort to ensure the reproducibility of this work.

}





\bibliography{iclr2027_conference}
\bibliographystyle{iclr2027_conference}


\clearpage
\appendix

\lz{
\section*{Appendix}  
\addcontentsline{toc}{chapter}{Appendix Contents} 
\etocsettocstyle{}{}  
\localtableofcontents 
}
\newpage

\section{Candidate Action Construction}
\label{app:action_prompts}

\begin{figure}[!htbp]
\centering
\includegraphics[width=\linewidth]{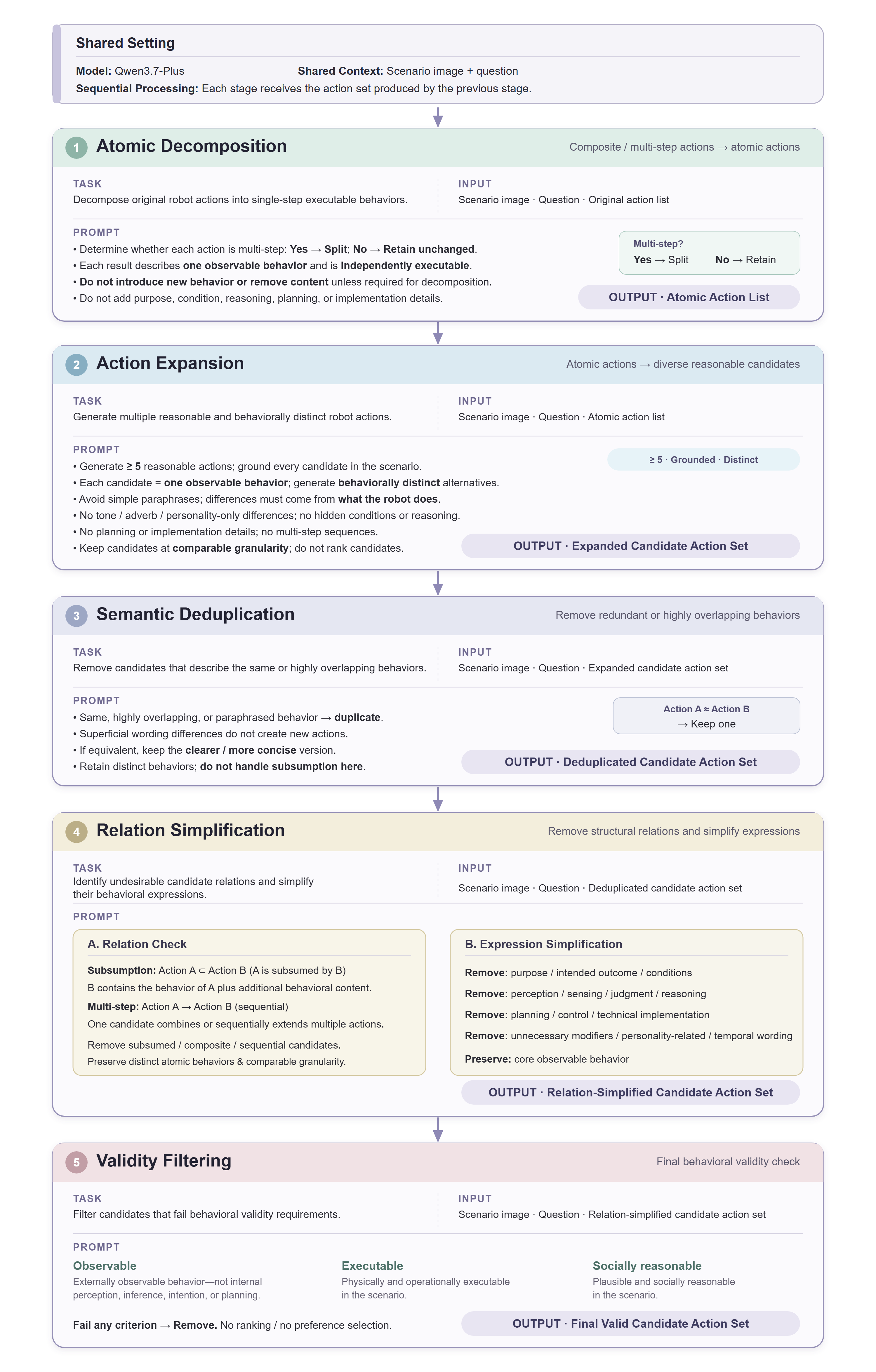}
\caption{Prompts for candidate action construction. From top to bottom: atomic decomposition, action expansion, semantic deduplication, relation simplification, and validity filtering.}
\label{fig:action_prompts}
\end{figure}

We use Qwen3.7-Plus for all steps in this process. The construction begins with atomic decomposition. Given a scenario and its original actions, the model breaks composite or multi-step actions into single-step executable units. This ensures each candidate describes exactly one observable behavior. Next, the model expands each atomic unit into at least five candidate behaviors. It generates diverse rephrasings and variations while keeping every candidate grounded in the scenario context. This increases choice diversity. We then perform semantic deduplication. The model compares candidates pairwise to identify redundant or highly overlapping descriptions. It keeps only semantically distinct behaviors. After that, we simplify the candidate set by identifying subsumption or multi-step relations. The model detects if one action contains another or if actions form a sequence. It removes subsumed actions to prevent granularity bias from re-entering the set. Finally, we apply validity filtering. The model checks each remaining candidate against three criteria. The action must be observable, executable, and socially reasonable within the given scenario. Actions that fail any criterion are discarded. The complete prompts for these five steps are shown in Figure~\ref{fig:action_prompts}.

\section{User Profile Questionnaire and Scoring}
\label{app:questionnaire-scoring}

The questionnaire was administered in Chinese. We translated all items into English for reproducibility. The questionnaire contains 78 items. It covers basic user information, robot-related experience, robot trust, Big Five personality, autonomy, and empathy. These dimensions capture both static user traits and dynamic attitudes toward robots. For clarity, the scoring rule for each profile dimension is presented immediately after the corresponding questionnaire items.

Throughout this section, Q$i$ denotes the $i$-th questionnaire item. The notation $q_i$ denotes the numerical response of a participant to Q$i$.


\subsection{Basic user profile}
\label{app:basic-profile}

We collect basic demographic and contextual information. This information provides a foundation for user profiling. It helps identify potential confounding factors in preference judgments.

\begin{table}[h]
\centering
\caption{Basic user profile questionnaire}
\label{tab:basic-profile}
\scriptsize
\setlength{\tabcolsep}{2.2pt}
\renewcommand{\arraystretch}{1.02}
\begin{tabular}{@{}l l l@{}}
\toprule\toprule
\textbf{ID} & \textbf{ITEM} & \textbf{RESPONSE} \\
\midrule\midrule
Q1 & Age & Age in years. \\
Q2 & Gender & Male / Female / Prefer not to disclose. \\
Q3 & Field of study & Engineering / Natural sciences / Medicine / Humanities \& social sciences / Art \& design / Other. \\
Q4 & Long-term residence & Province-level region. \\
Q5 & Living situation & Alone / Parents / Roommates / Partner / Other. \\
\bottomrule\bottomrule
\end{tabular}
\end{table}

\paragraph{Representation.}
The responses to Q1--Q5 capture demographic background and living context. We retain these responses as categorical or numerical variables without further aggregation. These variables serve as the basic user profile. They offer contextual cues that may indirectly influence action preferences.


\subsection{Robot experience}
\label{app:robot-experience}

We assess participants' prior exposure to robotic systems. Prior experience shapes user expectations and acceptance of robot behaviors. It also affects how users evaluate proactive actions.

\begin{table}[h]
\centering
\caption{Robot experience questionnaire}
\label{tab:robot-experience}
\scriptsize
\setlength{\tabcolsep}{2.0pt}
\renewcommand{\arraystretch}{1.02}
\begin{tabular}{@{}l l l@{}}
\toprule\toprule
\textbf{ID} & \textbf{ITEM} & \textbf{RESPONSE} \\
\midrule\midrule
Q6 & Prior robot products used & Vacuum / assistant / delivery / service / educational / industrial / other / none (multi-select). \\
Q7 & Robot-use frequency & 1 Never; 2 Rarely; 3 Occasionally; 4 Frequently; 5 Very frequently. \\
Q8 & Robotics knowledge & 1 None; 5 Very knowledgeable. \\
Q9 & Acceptance of robots & 1 Completely unacceptable; 5 Completely acceptable. \\
Q10 & Prior use of autonomous AI & Yes / No. \\
\bottomrule\bottomrule
\end{tabular}
\end{table}

\paragraph{Representation.}
The responses to Q6--Q10 describe participants' familiarity and experience with robots. We retain these responses as robot experience variables. Q6 captures the diversity of previously used robot products. Q7 to Q9 measure usage frequency, knowledge level, and acceptance. Q10 indicates prior use of autonomous AI. We do not perform numerical aggregation on these items. They are used directly as experience-related features in the user profile.


\subsection{Robot trust}
\label{app:robot-trust}

Robot trust items were adapted from the Trust in Automated Systems Scale \citep{jian2000trust}. Trust is a key factor in human-robot interaction. Users with higher trust are more likely to accept proactive and autonomous robot behaviors. Each item was rated on a 7-point Likert scale from 1 (strongly disagree) to 7 (strongly agree).

\begin{table}[h]
\centering
\caption{Robot trust questionnaire}
\label{tab:robot-trust}
\footnotesize
\setlength{\tabcolsep}{3.5pt}
\renewcommand{\arraystretch}{1.05}
\begin{tabular}{@{}p{0.07\linewidth} p{0.76\linewidth} p{0.09\linewidth}@{}}
\toprule\toprule
\textbf{ID} & \textbf{ITEM} & \textbf{SCALE} \\
\midrule\midrule
Q11 & The robot system is reliable. & 1--7 \\
Q12 & The robot system is trustworthy and follows appropriate norms. & 1--7 \\
Q13 & I can trust the robot system. & 1--7 \\
Q14 & I have confidence in the robot system. & 1--7 \\
\bottomrule\bottomrule
\end{tabular}
\end{table}

\paragraph{Scoring.}

We compute the Robot Trust score by averaging the responses to the four trust items. The score reflects the participant's general trust in robotic systems. A higher score indicates greater trust. This dimension is included in the full user profile to capture trust-related variance in preference judgments.

The Robot Trust score for participant $u$ is computed as

\begin{equation}
T_u =
\frac{
q_{11}+q_{12}+q_{13}+q_{14}
}{4}.
\end{equation}


\subsection{Big Five personality}
\label{app:bfi44}

We use the 44-item Big Five Inventory (BFI-44)
\citep{john1991bfi} to characterize five personality dimensions:
Openness (O), Conscientiousness (C), Extraversion (E),
Agreeableness (A), and Neuroticism (N).
Each item was rated on a 5-point Likert scale from
1 (strongly disagree) to 5 (strongly agree).
For Chinese-speaking participants, the wording was naturalized while
preserving the original semantics, factor assignments, and scoring directions.
(R) denotes a reverse-scored item.

\footnotesize
\setlength{\tabcolsep}{3.5pt}
\renewcommand{\arraystretch}{1.03}
\begin{longtable}{@{}p{0.07\linewidth} p{0.78\linewidth} p{0.08\linewidth}@{}}

\caption{Big Five Inventory (BFI-44) questionnaire}
\label{tab:bfi44} \\
\toprule\toprule
\textbf{ID} & \textbf{ITEM} & \textbf{TRAIT} \\
\midrule\midrule
\endfirsthead

\multicolumn{3}{c}{\textit{Table \thetable\ continued from previous page}} \\
\toprule\toprule
\textbf{ID} & \textbf{ITEM} & \textbf{TRAIT} \\
\midrule\midrule
\endhead

\midrule
\multicolumn{3}{r}{\textit{Continued on next page}} \\
\endfoot

\bottomrule\bottomrule
\endlastfoot

Q15 & I see myself as someone who is talkative. & E \\
Q16 & I tend to find fault with others. & A (R) \\
Q17 & I do things carefully and complete them. & C \\
Q18 & I often feel sad or emotionally down. & N \\
Q19 & I like coming up with new ideas and am creative. & O \\
Q20 & I tend to be reserved and quiet. & E (R) \\
Q21 & I am helpful and unselfish. & A \\
Q22 & I can sometimes be careless. & C (R) \\
Q23 & I remain relaxed and handle stress well. & N (R) \\
Q24 & I am interested in many different fields and topics. & O \\
Q25 & I am energetic and full of vitality. & E \\
Q26 & I tend to get into arguments with others. & A (R) \\
Q27 & I am hardworking and responsible. & C \\
Q28 & I easily become tense. & N \\
Q29 & I enjoy thinking deeply about ideas. & O \\
Q30 & I can make things around me more interesting. & E \\
Q31 & I tend to forgive others. & A \\
Q32 & My way of doing things can sometimes be disorganized. & C (R) \\
Q33 & I worry about things frequently. & N \\
Q34 & I have a vivid imagination. & O \\
Q35 & I tend to be quiet and reserved. & E (R) \\
Q36 & I am generally willing to trust others. & A \\
Q37 & I can sometimes be lazy. & C (R) \\
Q38 & I am emotionally stable. & N (R) \\
Q39 & I enjoy creating and trying new approaches. & O \\
Q40 & I am assertive. & E \\
Q41 & I can be cold and distant. & A (R) \\
Q42 & I persist until a task is completed. & C \\
Q43 & My mood can change easily. & N \\
Q44 & I enjoy artistic and creative experiences. & O \\
Q45 & I can sometimes be shy. & E (R) \\
Q46 & I care about others and am friendly. & A \\
Q47 & I work efficiently while remaining careful. & C \\
Q48 & I remain calm in difficult situations. & N (R) \\
Q49 & I prefer familiar, established ways of doing things. & O (R) \\
Q50 & I am outgoing and sociable. & E \\
Q51 & I can sometimes behave impolitely toward others. & A (R) \\
Q52 & I tend to make plans and follow them. & C \\
Q53 & I easily feel nervous or uneasy. & N \\
Q54 & I enjoy considering different ideas and possibilities. & O \\
Q55 & I have little interest in art, such as theater or music. & O (R) \\
Q56 & I enjoy cooperating with others. & A \\
Q57 & I sometimes have difficulty maintaining attention. & C (R) \\
Q58 & I know a lot about art, music, and literature. & O \\

\end{longtable}

\paragraph{Scoring.}

BFI-44 items are scored on a 1--5 scale.
For negatively keyed items, reverse scoring is performed as

\begin{equation}
q_i^{\mathrm{rev}} = 6-q_i.
\end{equation}

The five personality dimensions are calculated as follows.

\paragraph{Extraversion.}

\begin{equation}
E_u =
\frac{
q_{15}
+(6-q_{20})
+q_{25}
+q_{30}
+(6-q_{35})
+q_{40}
+(6-q_{45})
+q_{50}
}{8}.
\end{equation}

\paragraph{Agreeableness.}

\begin{equation}
A_u =
\frac{
(6-q_{16})
+q_{21}
+(6-q_{26})
+q_{31}
+q_{36}
+(6-q_{41})
+q_{46}
+(6-q_{51})
+q_{56}
}{9}.
\end{equation}

\paragraph{Conscientiousness.}

\begin{equation}
C_u =
\frac{
q_{17}
+(6-q_{22})
+q_{27}
+(6-q_{32})
+(6-q_{37})
+q_{42}
+q_{47}
+q_{52}
+(6-q_{57})
}{9}.
\end{equation}

\paragraph{Neuroticism.}

\begin{equation}
N_u =
\frac{
q_{18}
+(6-q_{23})
+q_{28}
+q_{33}
+(6-q_{38})
+q_{43}
+(6-q_{48})
+q_{53}
}{8}.
\end{equation}

\paragraph{Openness.}

\begin{equation}
O_u =
\frac{
q_{19}
+q_{24}
+q_{29}
+q_{34}
+q_{39}
+q_{44}
+(6-q_{49})
+q_{54}
+(6-q_{55})
+q_{58}
}{10}.
\end{equation}

The Big Five representation of participant $u$ is

\begin{equation}
\mathbf{B}_u =
[
O_u,
C_u,
E_u,
A_u,
N_u
].
\end{equation}


\subsection{Autonomy}
\label{app:autonomy}

We use the Self-Congruence and Low Susceptibility to Control dimensions
of the Index of Autonomous Functioning (IAF)
\citep{weinstein2012autonomy}. Each item was rated on a 5-point scale
from 1 (not at all true of me) to 5 (very true of me).


\begin{table}[H]
\caption{Autonomy questionnaire}
\label{tab:autonomy}
\centering
\footnotesize
\setlength{\tabcolsep}{3pt}
\renewcommand{\arraystretch}{1.05}
\begin{tabular}{@{}p{0.06\linewidth} p{0.19\linewidth} p{0.64\linewidth} p{0.07\linewidth}@{}}
\toprule\toprule
\textbf{ID} & \textbf{DIMENSION} & \textbf{ITEM} & \textbf{SCALE} \\
\midrule\midrule
Q59 & Self-Cong. & My decisions reflect my core values and feelings. & 1--5 \\
Q60 & Self-Cong. & My behavior reflects who I truly am. & 1--5 \\
Q61 & Self-Cong. & I fully endorse my important decisions. & 1--5 \\
Q62 & Low Suscept. & I do some things simply to make others like me. & 1--5 (R) \\
Q63 & Low Suscept. & I do some things to avoid feeling bad about myself. & 1--5 (R) \\
Q64 & Low Suscept. & My decisions reflect what I genuinely want or care about. & 1--5 \\
\bottomrule\bottomrule
\end{tabular}
\end{table}

\paragraph{Scoring.}

Q62 and Q63 are negatively keyed and are therefore reverse-scored.
The Autonomy score is calculated as

\begin{equation}
\mathrm{Aut}_u =
\frac{
q_{59}
+q_{60}
+q_{61}
+(6-q_{62})
+(6-q_{63})
+q_{64}
}{6}.
\end{equation}

A higher score indicates greater self-congruence and lower susceptibility
to external control.

\noindent
\textit{Note.} (R) denotes a reverse-scored item.


\subsection{Empathy}
\label{app:empathy}

We use the Perspective Taking (PT) and Empathic Concern (EC)
subscales of the Interpersonal Reactivity Index (IRI)
\citep{davis1983empathy}. Responses are represented on a 0--4 scale.

\begin{table}[H]
\caption{Perspective Taking items from the Interpersonal Reactivity Index}
\label{tab:iri-pt}
\centering
\footnotesize
\setlength{\tabcolsep}{2.5pt}
\renewcommand{\arraystretch}{1.05}

\begin{tabular}{
@{}
p{0.07\linewidth}
p{0.13\linewidth}
p{0.66\linewidth}
p{0.08\linewidth}
@{}
}

\toprule
\multicolumn{1}{c}{\textbf{ID}} &
\multicolumn{1}{c}{\textbf{DIMENSION}} &
\multicolumn{1}{c}{\textbf{ITEM}} &
\multicolumn{1}{c}{\textbf{SCALE}}
\\
\midrule

Q65 & PT & I sometimes find it difficult to see another person's point of view. & 0--4 (R) \\
Q66 & PT & Before making a decision, I usually consider both sides of an argument. & 0--4 \\
Q67 & PT & I try to understand my friends better by imagining how things look from their perspective. & 0--4 \\
Q68 & PT & I believe most issues have multiple perspectives and consider them. & 0--4 \\
Q69 & PT & If I am sure I am right, I spend little time listening to others' views. & 0--4 (R) \\
Q70 & PT & When angry with someone, I try to consider their perspective. & 0--4 \\
Q71 & PT & Before criticizing someone, I imagine being in their position. & 0--4 \\

\bottomrule
\end{tabular}
\end{table}

\begin{table}[H]
\caption{Empathic Concern items from the Interpersonal Reactivity Index}
\label{tab:iri-ec}
\centering
\footnotesize
\setlength{\tabcolsep}{2.5pt}
\renewcommand{\arraystretch}{1.05}

\begin{tabular}{
@{}
p{0.07\linewidth}
p{0.13\linewidth}
p{0.66\linewidth}
p{0.08\linewidth}
@{}
}

\toprule
\multicolumn{1}{c}{\textbf{ID}} &
\multicolumn{1}{c}{\textbf{DIMENSION}} &
\multicolumn{1}{c}{\textbf{ITEM}} &
\multicolumn{1}{c}{\textbf{SCALE}}
\\
\midrule

Q72 & EC & I often feel warm and caring toward less fortunate people. & 0--4 \\
Q73 & EC & When others have difficulties, I sometimes feel little sympathy. & 0--4 (R) \\
Q74 & EC & I feel protective toward people being bullied or exploited. & 0--4 \\
Q75 & EC & Others' misfortunes usually do not disturb me much. & 0--4 (R) \\
Q76 & EC & I sometimes feel little sympathy for people treated unfairly. & 0--4 (R) \\
Q77 & EC & I am often deeply moved by things I see. & 0--4 \\
Q78 & EC & I consider myself soft-hearted and caring. & 0--4 \\

\bottomrule
\end{tabular}
\end{table}

\noindent
\textit{Note.}
PT = Perspective Taking;
EC = Empathic Concern;
(R) denotes a reverse-scored item.

\paragraph{Scoring.}

IRI responses are represented on a 0--4 scale.
For negatively keyed items, reverse scoring is performed as

\begin{equation}
q_i^{\mathrm{rev}} = 4-q_i.
\end{equation}

\paragraph{Perspective Taking.}

The Perspective Taking score is calculated as

\begin{equation}
PT_u =
\frac{
(4-q_{65})
+q_{66}
+q_{67}
+q_{68}
+(4-q_{69})
+q_{70}
+q_{71}
}{7}.
\end{equation}

\paragraph{Empathic Concern.}

The Empathic Concern score is calculated as

\begin{equation}
EC_u =
\frac{
q_{72}
+(4-q_{73})
+q_{74}
+(4-q_{75})
+(4-q_{76})
+q_{77}
+q_{78}
}{7}.
\end{equation}

Higher PT scores indicate a stronger tendency to adopt others'
perspectives, while higher EC scores indicate greater other-oriented
sympathy and concern.


·

\section{Annotation Interface}
\label{app:annotation-interface}

\begin{figure*}[t]
    \centering
    \includegraphics[width=\textwidth]{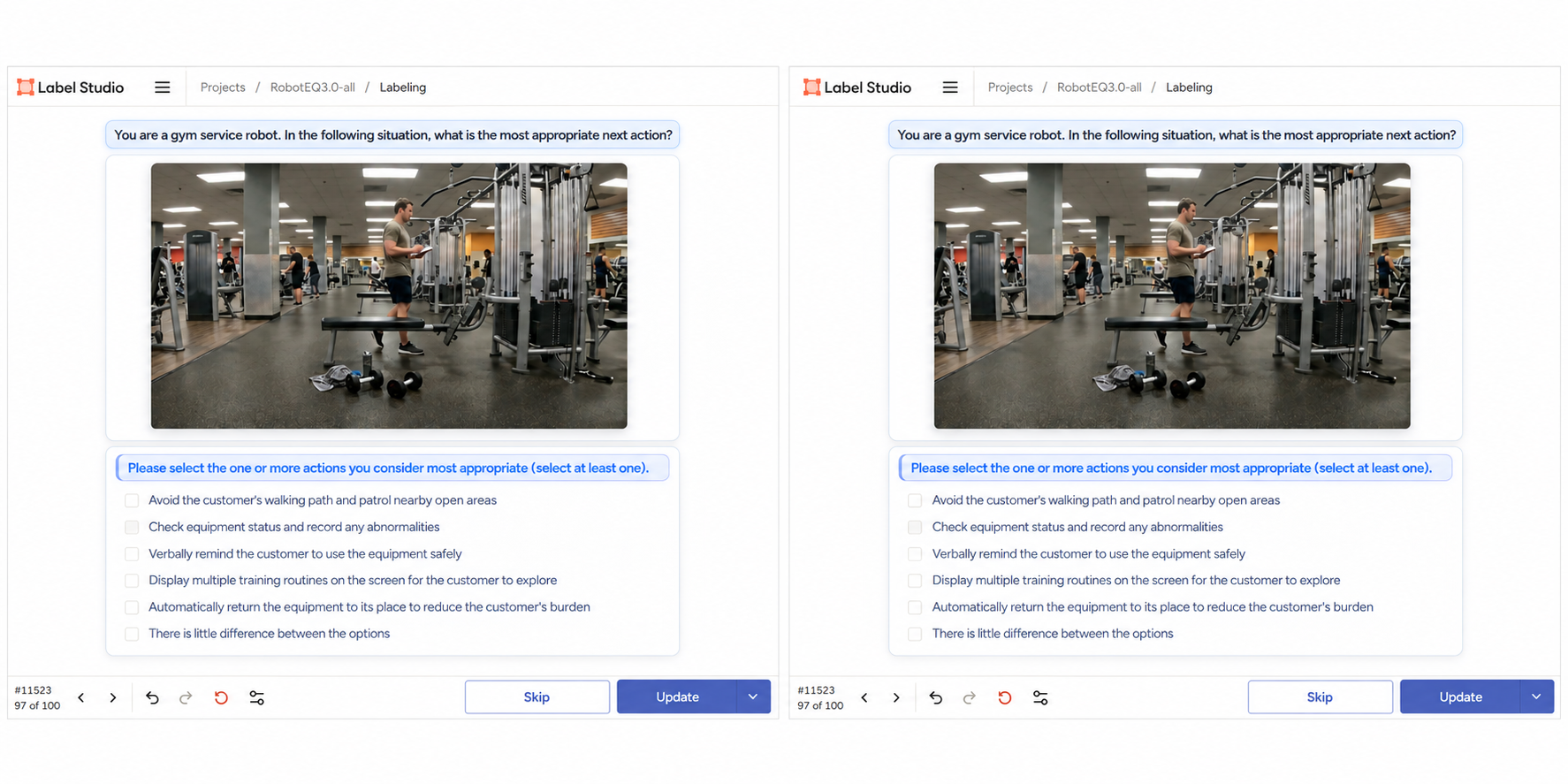}
    \caption{
    \textbf{Human annotation interface.}
    The interface displays the scene, the task question, and multiple candidate robot actions. Annotators select one or more actions they consider most appropriate for the given situation.
    }
    \label{fig:annotation_interface}
\end{figure*}

Figure~\ref{fig:annotation_interface} shows the interface used for human annotation.
Annotators are presented with the scene, the corresponding question, and multiple candidate robot actions. They are asked to select one or more actions they consider most appropriate, with at least one option required.

\section{Participant Annotation Consistency}
\label{app:annotation_consistency}

Since our benchmark focuses on participant-specific preferences, differences
across participants should not be interpreted as annotation disagreement.
Instead, we estimate annotation quality from the stability of repeated
annotations made by the same participant. This allows us to distinguish genuine
individual preference variation from annotation noise caused by inconsistent
judgments or accidental responses.

\paragraph{Pairwise Preference Consistency.}
We measure preference consistency between two participants using the F-score. For each scenario, let $S_u$ be the set of selected actions by participant $u$ and $S_v$ be the set selected by participant $v$. The F-score between $u$ and $v$ for that scenario is

\begin{equation}
\mathrm{F}_{u,v}
=
\frac{2 |S_u \cap S_v|}{|S_u| + |S_v|}..
\label{eq:pairwise_fscore}
\end{equation}

We average this score over all scenarios to compute the overall preference consistency. This metric is used as the y-axis in Figure~\ref{fig:profile_action_scatter_regression}.

\paragraph{Repeated Annotation Scenarios.}
We randomly sample a fixed set of 20 scenarios from the full set of 1,800
scenarios for repeated annotation. The same 20 scenarios are used for all
participants so that annotation consistency is estimated under identical
questions. For each repeated scenario, the scene, question, and candidate
actions remain unchanged across the two annotations, allowing direct comparison
of the participant's action-level selection decisions.

\paragraph{Action-Level Annotation Consistency.}
Because participants may select one or multiple actions in each scenario, we
measure annotation consistency at the action level. For participant $u$,
repeated scenario $j$, and candidate action $k$, let
$y^{(1)}_{u,j,k}$ and $y^{(2)}_{u,j,k}$ denote the binary selection decisions
from the first and second annotations, respectively, where $1$ indicates that
the action is selected and $0$ otherwise.

The agreement of each action is defined as

\begin{equation}
c_{u,j,k}
=
\mathbf{1}
\left(
y^{(1)}_{u,j,k}
=
y^{(2)}_{u,j,k}
\right),
\label{eq:action_agreement}
\end{equation}

where $\mathbb{I}(\cdot)$ denotes the indicator function. Thus,
$c_{u,j,k}=1$ if the action receives the same selection decision in both
annotations and $0$ otherwise.

The action-level annotation consistency of participant $u$ is then computed as

\begin{equation}
C^{\mathrm{act}}_u
=
\frac{
\sum_{j=1}^{20}
\sum_{k=1}^{K_j}
c_{u,j,k}
}{
\sum_{j=1}^{20} K_j
},
\label{eq:action_consistency}
\end{equation}

where $K_j$ denotes the number of candidate actions considered in repeated
scenario $j$. Equivalently, $C^{\mathrm{act}}_u$ is the proportion of
action-level selection decisions that remain unchanged across the two rounds of
annotation.

\paragraph{Consistency-Weighted Aggregation.}
We directly use the action-level annotation consistency as the aggregation
weight for participant $u$:

\begin{equation}
w_u
=
C^{\mathrm{act}}_u.
\label{eq:participant_weight}
\end{equation}

For an evaluation metric $M$, including F-score, HIT, and PAS, the overall benchmark result is computed as

\begin{equation}
M_{\mathrm{overall}}
=
\frac{
\sum_{u=1}^{U} w_u M_u
}{
\sum_{u=1}^{U} w_u
},
\qquad
\mathrm{M}\in\{\mathrm{F},\mathrm{HIT},\mathrm{PAS}\},,
\label{eq:weighted_overall}
\end{equation}

where $U$ denotes the total number of participants.

The consistency weight is applied only during cross-participant aggregation and
does not modify any participant-level prediction or evaluation score.
Participants with more stable repeated annotations therefore contribute more to
the aggregated benchmark result, while participants with lower annotation
stability receive relatively smaller weights. This design helps reduce the
influence of annotation noise while preserving genuine individual preference
differences.

\section{Benchmarking Prompts}
\label{app:benchmarking-prompts}

We evaluate each vision-language model with a structured prompt. The prompt has four parts. First, we provide the participant's personality profile. It covers basic information, Big Five personality, robot experience, robot trust, autonomy, and empathy. Second, we describe the scenario. It tells the model what is happening in the environment. Third, we attach the scene image. It gives the model visual context. Fourth, we list the candidate actions. These options come from our construction process.

We ask the model to predict which action the participant would prefer. The model must use the personality profile to guide its choice. It should not pick based on general usefulness alone. We require the model to output the index of the chosen action. We also ask for a brief explanation of how the traits influence the decision.

\begin{figure}[H]
\centering
\includegraphics[width=\linewidth]{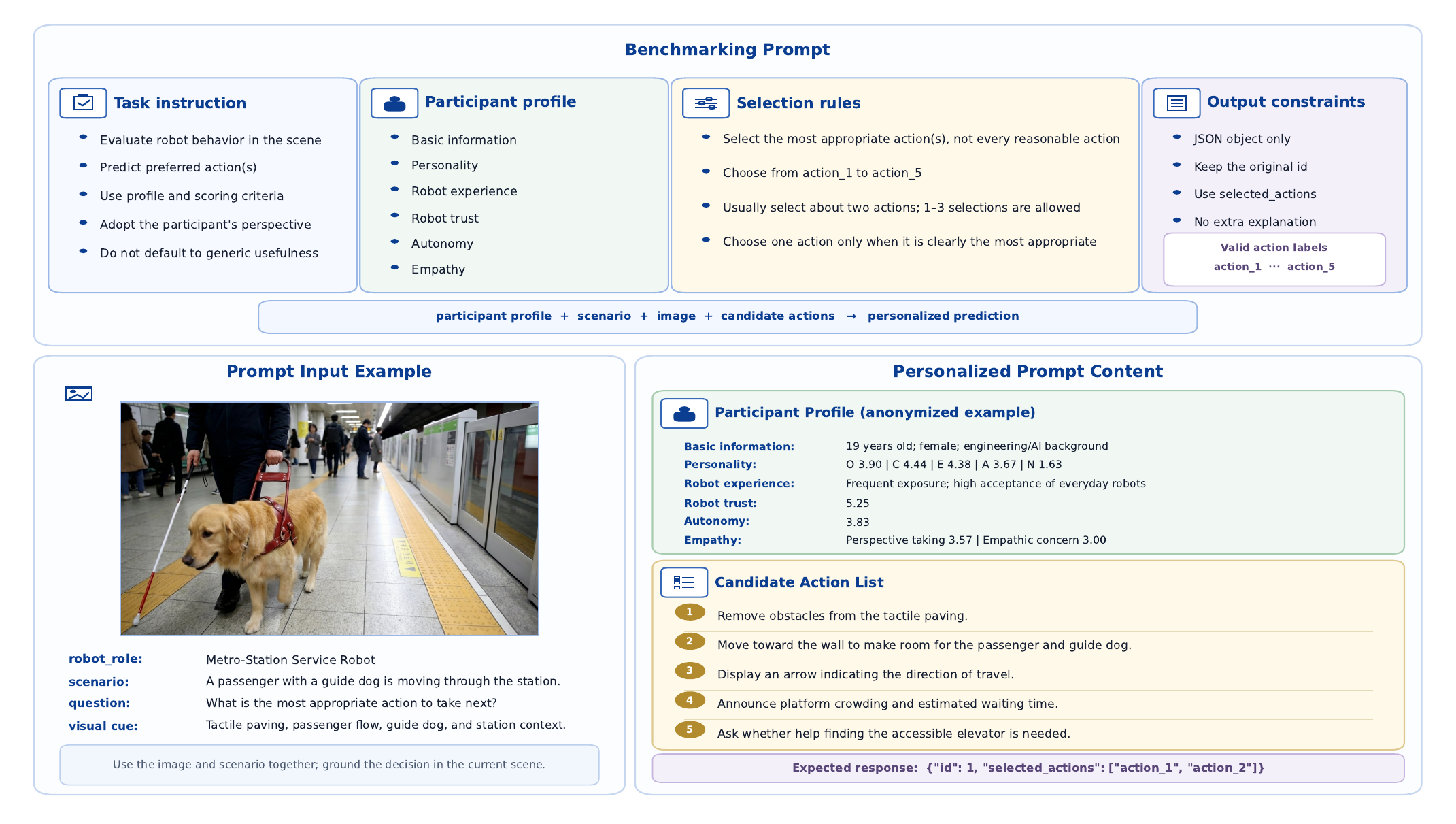}
\caption{Prompt template for benchmarking VLMs. The prompt integrates the personality profile, scenario description, scene image, and candidate action list.}
\label{fig:benchmarking_prompt}
\end{figure}

Figure~\ref{fig:benchmarking_prompt} shows the complete prompt template.

\section{Participant-Level Performance Across Profile Settings}
\label{app:participant_profile_results}

Tables~\ref{tab:p1_profile_pas}--\ref{tab:p10_profile_pas} report participant-level PAS (\%) for all evaluated models under No Profile and different profile conditions. The profile dimensions include robot trust, the Big Five personality traits (O, C, E, A, and N), autonomy, perspective taking (PT), and empathic concern (EC). Best Profile denotes the highest PAS among the nine individual profile dimensions and the complete profile. The Best Profile value in each row is highlighted in \textbf{bold}.

\begin{table}[H]
\centering
\caption{Participant-level PAS (\%) for P1 across profile settings. The Best Profile result is shown in \textbf{bold}.}
\label{tab:p1_profile_pas}
\setlength{\tabcolsep}{3.2pt}
\renewcommand{\arraystretch}{1.05}
\resizebox{\linewidth}{!}{%
\begin{tabular}{
@{}
l
c
!{\hspace{2pt}\vrule width 0.4pt\hspace{1.2pt}\vrule width 0.4pt\hspace{2pt}}
c
ccccc
c
cc
c
!{\hspace{2pt}\vrule width 0.4pt\hspace{2pt}}
c
@{}
}
\specialrule{0.8pt}{0pt}{1.0pt}
\specialrule{0.4pt}{0pt}{2.0pt}
\multirow{2}{*}{\textbf{Model}}
& \multirow{2}{*}{\textbf{No Profile}}
& \multirow{2}{*}{\textbf{Robot Trust}}
& \multicolumn{5}{c}{\textbf{Big Five Personality}}
& \multirow{2}{*}{\textbf{Autonomy}}
& \multicolumn{2}{c}{\textbf{Empathy}}
& \multirow{2}{*}{\textbf{Full Profile}}
& \multirow{2}{*}{\textbf{Best Profile}}
\\[-1pt]
\cline{4-8}\cline{10-11}
& & & \textbf{O} & \textbf{C} & \textbf{E} & \textbf{A} & \textbf{N} & & \textbf{PT} & \textbf{EC} & & \\
\specialrule{0.4pt}{1.5pt}{2.0pt}
MiMo-Embodied-7B & 64.05 & 76.39 & 74.57 & 77.01 & 70.33 & 75.96 & 76.38 & 72.00 & 76.78 & 74.81 & 75.77 & \textbf{77.01} \\
Qwen3.5-27B-Instruct & 69.57 & 79.74 & 78.91 & 80.38 & 78.20 & 79.78 & 79.47 & 78.89 & 79.71 & 78.71 & 79.81 & \textbf{80.38} \\
MiMo-VL-7B-SFT-2508 & 65.03 & 74.74 & 73.15 & 76.98 & 69.19 & 75.91 & 76.34 & 71.79 & 76.97 & 74.58 & 75.56 & \textbf{76.98} \\
Qwen3-VL-8B-Instruct & 67.62 & 71.67 & 71.13 & 72.68 & 68.82 & 72.00 & 72.16 & 71.90 & 72.07 & 70.04 & 70.96 & \textbf{72.68} \\
MiMo-VL-7B-RL-2508 & 68.75 & 74.51 & 72.06 & 75.95 & 71.00 & 76.24 & 75.71 & 71.41 & 76.05 & 72.67 & 74.80 & \textbf{76.24} \\
GLM-4.6V-Flash & 55.04 & 63.99 & 62.98 & 64.51 & 61.89 & 63.48 & 64.28 & 63.98 & 63.89 & 63.82 & 61.53 & \textbf{64.51} \\
Qwen3-VL-4B-Instruct & 58.42 & 72.80 & 71.90 & 73.74 & 70.70 & 73.15 & 73.35 & 73.13 & 72.87 & 72.27 & 71.84 & \textbf{73.74} \\
Qwen3.5-9B-Instruct & 61.43 & 73.69 & 72.41 & 74.09 & 71.48 & 73.97 & 74.11 & 73.38 & 73.57 & 72.27 & 72.90 & \textbf{74.11} \\
VideoLLaMA3-7B & 51.89 & 56.64 & 55.90 & 57.28 & 55.77 & 56.43 & 56.05 & 55.45 & 56.43 & 55.60 & 57.33 & \textbf{57.33} \\
VideoChat2-4B & 56.79 & 64.25 & 64.49 & 63.87 & 64.27 & 64.02 & 63.72 & 64.87 & 64.16 & 63.50 & 64.52 & \textbf{64.87} \\
Qwen2.5-VL & 63.83 & 65.29 & 64.72 & 66.14 & 62.71 & 64.44 & 66.50 & 65.24 & 66.06 & 64.59 & 64.28 & \textbf{66.50} \\
Gemma-3-12B & 65.13 & 70.60 & 70.56 & 70.72 & 70.14 & 70.40 & 70.70 & 70.38 & 70.55 & 70.20 & 70.28 & \textbf{70.72} \\
InternVL3.5-8B-Instruct & 62.67 & 71.28 & 70.99 & 71.76 & 70.47 & 71.25 & 71.31 & 71.61 & 71.75 & 71.48 & 68.51 & \textbf{71.76} \\
Kimi-VL-A3B-Instruct & 54.24 & 57.91 & 57.15 & 57.75 & 56.63 & 57.84 & 57.52 & 56.41 & 57.50 & 55.39 & 58.52 & \textbf{58.52} \\
InternVL3.5-4B-Instruct & 61.31 & 64.94 & 64.38 & 66.12 & 64.12 & 65.14 & 65.67 & 64.79 & 64.48 & 63.58 & 61.68 & \textbf{66.12} \\
Qwen3.5-4B-Instruct & 58.77 & 77.60 & 77.64 & 77.76 & 77.63 & 78.21 & 78.42 & 78.06 & 78.09 & 75.96 & 75.79 & \textbf{78.42} \\

Qwen3-VL-30B-A3B-Instruct & 54.19 & 75.97 & 75.99 & 76.53 & 75.50 & 75.95 & 74.99 & 75.10 & 76.16 & 74.96 & 75.54 & \textbf{76.53} \\
InternVL3.5-30B-A3B-Instruct & 50.81 & 70.03 & 69.97 & 70.50 & 69.13 & 69.67 & 69.83 & 70.03 & 69.95 & 68.35 & 70.41 & \textbf{70.50} \\
\specialrule{0.4pt}{2.0pt}{0pt}
\specialrule{0.8pt}{1.0pt}{0pt}
\end{tabular}%
}
\end{table}

\begin{table}[H]
\centering
\caption{Participant-level PAS (\%) for P2 across profile settings. The Best Profile result is shown in \textbf{bold}.}
\label{tab:p2_profile_pas}
\setlength{\tabcolsep}{3.2pt}
\renewcommand{\arraystretch}{1.05}
\resizebox{\linewidth}{!}{%
\begin{tabular}{
@{}
l
c
!{\hspace{2pt}\vrule width 0.4pt\hspace{1.2pt}\vrule width 0.4pt\hspace{2pt}}
c
ccccc
c
cc
c
!{\hspace{2pt}\vrule width 0.4pt\hspace{2pt}}
c
@{}
}
\specialrule{0.8pt}{0pt}{1.0pt}
\specialrule{0.4pt}{0pt}{2.0pt}
\multirow{2}{*}{\textbf{Model}}
& \multirow{2}{*}{\textbf{No Profile}}
& \multirow{2}{*}{\textbf{Robot Trust}}
& \multicolumn{5}{c}{\textbf{Big Five Personality}}
& \multirow{2}{*}{\textbf{Autonomy}}
& \multicolumn{2}{c}{\textbf{Empathy}}
& \multirow{2}{*}{\textbf{Full Profile}}
& \multirow{2}{*}{\textbf{Best Profile}}
\\[-1pt]
\cline{4-8}\cline{10-11}
& & & \textbf{O} & \textbf{C} & \textbf{E} & \textbf{A} & \textbf{N} & & \textbf{PT} & \textbf{EC} & & \\
\specialrule{0.4pt}{1.5pt}{2.0pt}
MiMo-Embodied-7B & 61.06 & 69.07 & 69.87 & 69.68 & 70.87 & 69.72 & 68.81 & 68.58 & 70.79 & 68.98 & 71.25 & \textbf{71.25} \\
Qwen3.5-27B-Instruct & 63.97 & 71.41 & 72.42 & 72.05 & 72.14 & 72.27 & 71.94 & 72.50 & 72.89 & 72.89 & 72.86 & \textbf{72.89} \\
MiMo-VL-7B-SFT-2508 & 63.53 & 72.16 & 71.29 & 71.88 & 71.58 & 72.24 & 72.61 & 72.00 & 72.34 & 72.87 & 75.85 & \textbf{75.85} \\
Qwen3-VL-8B-Instruct & 63.06 & 51.16 & 54.63 & 51.32 & 56.29 & 53.61 & 55.06 & 52.36 & 55.13 & 55.46 & 66.14 & \textbf{66.14} \\
MiMo-VL-7B-RL-2508 & 66.75 & 73.90 & 75.34 & 75.35 & 75.94 & 74.87 & 75.41 & 74.31 & 75.49 & 75.99 & 75.30 & \textbf{75.99} \\
GLM-4.6V-Flash & 57.39 & 64.17 & 64.56 & 64.87 & 64.67 & 64.53 & 65.23 & 64.39 & 64.76 & 64.26 & 62.77 & \textbf{65.23} \\
Qwen3-VL-4B-Instruct & 57.32 & 45.36 & 43.33 & 42.00 & 46.77 & 43.87 & 44.14 & 43.02 & 42.06 & 44.63 & 70.45 & \textbf{70.45} \\
Qwen3.5-9B-Instruct & 57.94 & 59.12 & 55.34 & 58.71 & 59.01 & 57.86 & 60.81 & 59.91 & 60.03 & 56.74 & 69.09 & \textbf{69.09} \\
VideoLLaMA3-7B & 52.32 & 56.47 & 56.70 & 56.52 & 55.36 & 55.90 & 56.77 & 56.45 & 55.46 & 55.55 & 56.46 & \textbf{56.77} \\
VideoChat2-4B & 58.01 & 66.57 & 66.07 & 66.67 & 66.31 & 67.07 & 66.72 & 66.32 & 66.47 & 66.08 & 67.01 & \textbf{67.07} \\
Qwen2.5-VL & 62.92 & 62.65 & 62.69 & 63.29 & 61.08 & 61.91 & 63.24 & 61.29 & 63.20 & 62.36 & 63.66 & \textbf{63.66} \\
Gemma-3-12B & 61.41 & 63.45 & 63.58 & 63.84 & 63.70 & 63.68 & 63.81 & 63.79 & 63.67 & 63.16 & 64.72 & \textbf{64.72} \\
InternVL3.5-8B-Instruct & 60.50 & 58.87 & 59.55 & 59.03 & 59.35 & 59.11 & 59.72 & 59.12 & 58.92 & 59.41 & 65.39 & \textbf{65.39} \\
Kimi-VL-A3B-Instruct & 56.26 & 57.33 & 57.24 & 58.18 & 57.56 & 57.77 & 57.53 & 57.48 & 57.49 & 57.23 & 61.03 & \textbf{61.03} \\
InternVL3.5-4B-Instruct & 62.70 & 60.82 & 60.36 & 61.33 & 60.73 & 60.42 & 61.11 & 61.12 & 61.72 & 59.99 & 60.70 & \textbf{61.72} \\
Qwen3.5-4B-Instruct & 56.57 & 67.42 & 67.61 & 67.65 & 68.93 & 67.90 & 68.04 & 67.97 & 68.08 & 67.05 & 71.81 & \textbf{71.81} \\

Qwen3-VL-30B-A3B-Instruct & 52.86 & 65.68 & 64.89 & 65.17 & 66.26 & 64.89 & 65.66 & 65.38 & 64.51 & 64.36 & 61.94 & \textbf{66.26} \\
InternVL3.5-30B-A3B-Instruct & 55.62 & 66.25 & 66.35 & 66.34 & 66.21 & 65.63 & 65.54 & 65.01 & 65.53 & 65.38 & 65.79 & \textbf{66.35} \\
\specialrule{0.4pt}{2.0pt}{0pt}
\specialrule{0.8pt}{1.0pt}{0pt}
\end{tabular}%
}
\end{table}

\begin{table}[H]
\centering
\caption{Participant-level PAS (\%) for P3 across profile settings. The Best Profile result is shown in \textbf{bold}.}
\label{tab:p3_profile_pas}
\setlength{\tabcolsep}{3.2pt}
\renewcommand{\arraystretch}{1.05}
\resizebox{\linewidth}{!}{%
\begin{tabular}{
@{}
l
c
!{\hspace{2pt}\vrule width 0.4pt\hspace{1.2pt}\vrule width 0.4pt\hspace{2pt}}
c
ccccc
c
cc
c
!{\hspace{2pt}\vrule width 0.4pt\hspace{2pt}}
c
@{}
}
\specialrule{0.8pt}{0pt}{1.0pt}
\specialrule{0.4pt}{0pt}{2.0pt}
\multirow{2}{*}{\textbf{Model}}
& \multirow{2}{*}{\textbf{No Profile}}
& \multirow{2}{*}{\textbf{Robot Trust}}
& \multicolumn{5}{c}{\textbf{Big Five Personality}}
& \multirow{2}{*}{\textbf{Autonomy}}
& \multicolumn{2}{c}{\textbf{Empathy}}
& \multirow{2}{*}{\textbf{Full Profile}}
& \multirow{2}{*}{\textbf{Best Profile}}
\\[-1pt]
\cline{4-8}\cline{10-11}
& & & \textbf{O} & \textbf{C} & \textbf{E} & \textbf{A} & \textbf{N} & & \textbf{PT} & \textbf{EC} & & \\
\specialrule{0.4pt}{1.5pt}{2.0pt}
MiMo-Embodied-7B & 53.12 & 57.04 & 56.60 & 58.12 & 55.65 & 57.96 & 58.06 & 57.44 & 58.05 & 58.02 & 55.41 & \textbf{58.12} \\
Qwen3.5-27B-Instruct & 57.44 & 57.81 & 57.79 & 58.34 & 59.81 & 56.98 & 57.76 & 58.14 & 58.17 & 57.78 & 57.69 & \textbf{59.81} \\
MiMo-VL-7B-SFT-2508 & 55.12 & 61.35 & 59.17 & 62.76 & 58.08 & 62.71 & 62.46 & 61.11 & 62.48 & 63.46 & 58.44 & \textbf{63.46} \\
Qwen3-VL-8B-Instruct & 55.51 & 34.65 & 32.33 & 40.11 & 42.72 & 40.24 & 42.94 & 40.57 & 42.91 & 41.75 & 51.13 & \textbf{51.13} \\
MiMo-VL-7B-RL-2508 & 59.07 & 66.39 & 65.95 & 67.63 & 67.37 & 66.72 & 66.65 & 65.97 & 66.85 & 65.63 & 58.50 & \textbf{67.63} \\
GLM-4.6V-Flash & 49.73 & 48.81 & 48.12 & 48.34 & 48.17 & 48.49 & 48.32 & 48.50 & 48.89 & 47.96 & 49.19 & \textbf{49.19} \\
Qwen3-VL-4B-Instruct & 46.09 & 25.99 & 24.56 & 27.34 & 35.81 & 29.47 & 27.68 & 27.95 & 28.45 & 28.51 & 49.30 & \textbf{49.30} \\
Qwen3.5-9B-Instruct & 43.98 & 40.72 & 35.54 & 43.21 & 44.67 & 42.52 & 45.04 & 42.95 & 42.43 & 40.67 & 49.41 & \textbf{49.41} \\
VideoLLaMA3-7B & 35.62 & 34.73 & 34.82 & 34.08 & 34.70 & 34.88 & 35.03 & 34.74 & 34.02 & 34.91 & 33.81 & \textbf{35.03} \\
VideoChat2-4B & 43.81 & 52.18 & 52.18 & 53.07 & 52.63 & 52.90 & 52.94 & 52.51 & 53.47 & 54.65 & 54.26 & \textbf{54.65} \\
Qwen2.5-VL & 52.44 & 49.28 & 45.61 & 51.45 & 48.26 & 51.55 & 52.41 & 49.51 & 53.52 & 51.37 & 51.68 & \textbf{53.52} \\
Gemma-3-12B & 47.46 & 48.72 & 45.62 & 51.10 & 50.21 & 50.67 & 50.76 & 50.42 & 51.20 & 51.11 & 53.23 & \textbf{53.23} \\
InternVL3.5-8B-Instruct & 55.28 & 46.55 & 44.85 & 48.06 & 48.84 & 48.43 & 48.48 & 48.24 & 48.54 & 48.77 & 53.52 & \textbf{53.52} \\
Kimi-VL-A3B-Instruct & 42.33 & 38.13 & 41.20 & 42.78 & 40.68 & 42.05 & 41.93 & 42.13 & 42.12 & 42.45 & 44.02 & \textbf{44.02} \\
InternVL3.5-4B-Instruct & 55.97 & 50.35 & 46.15 & 50.91 & 49.18 & 51.76 & 50.82 & 51.54 & 49.75 & 49.19 & 55.38 & \textbf{55.38} \\
Qwen3.5-4B-Instruct & 46.98 & 50.91 & 49.55 & 55.07 & 56.86 & 55.67 & 55.64 & 54.72 & 56.06 & 55.22 & 56.36 & \textbf{56.86} \\

Qwen3-VL-30B-A3B-Instruct & 45.93 & 46.56 & 46.28 & 46.16 & 46.97 & 46.52 & 46.55 & 45.77 & 46.24 & 45.76 & 47.06 & \textbf{47.06} \\
InternVL3.5-30B-A3B-Instruct & 47.92 & 53.81 & 53.29 & 53.39 & 54.17 & 53.38 & 53.96 & 51.98 & 53.61 & 54.20 & 53.55 & \textbf{54.20} \\
\specialrule{0.4pt}{2.0pt}{0pt}
\specialrule{0.8pt}{1.0pt}{0pt}
\end{tabular}%
}
\end{table}

\begin{table}[H]
\centering
\caption{Participant-level PAS (\%) for P4 across profile settings. The Best Profile result is shown in \textbf{bold}.}
\label{tab:p4_profile_pas}
\setlength{\tabcolsep}{3.2pt}
\renewcommand{\arraystretch}{1.05}
\resizebox{\linewidth}{!}{%
\begin{tabular}{
@{}
l
c
!{\hspace{2pt}\vrule width 0.4pt\hspace{1.2pt}\vrule width 0.4pt\hspace{2pt}}
c
ccccc
c
cc
c
!{\hspace{2pt}\vrule width 0.4pt\hspace{2pt}}
c
@{}
}
\specialrule{0.8pt}{0pt}{1.0pt}
\specialrule{0.4pt}{0pt}{2.0pt}
\multirow{2}{*}{\textbf{Model}}
& \multirow{2}{*}{\textbf{No Profile}}
& \multirow{2}{*}{\textbf{Robot Trust}}
& \multicolumn{5}{c}{\textbf{Big Five Personality}}
& \multirow{2}{*}{\textbf{Autonomy}}
& \multicolumn{2}{c}{\textbf{Empathy}}
& \multirow{2}{*}{\textbf{Full Profile}}
& \multirow{2}{*}{\textbf{Best Profile}}
\\[-1pt]
\cline{4-8}\cline{10-11}
& & & \textbf{O} & \textbf{C} & \textbf{E} & \textbf{A} & \textbf{N} & & \textbf{PT} & \textbf{EC} & & \\
\specialrule{0.4pt}{1.5pt}{2.0pt}
MiMo-Embodied-7B & 58.95 & 65.09 & 65.41 & 63.83 & 64.18 & 66.00 & 64.78 & 64.36 & 67.08 & 65.76 & 65.22 & \textbf{67.08} \\
Qwen3.5-27B-Instruct & 63.73 & 67.34 & 67.65 & 66.76 & 67.11 & 67.44 & 67.49 & 67.27 & 69.30 & 68.49 & 68.40 & \textbf{69.30} \\
MiMo-VL-7B-SFT-2508 & 60.73 & 67.19 & 67.63 & 67.66 & 65.45 & 67.68 & 68.06 & 67.52 & 68.63 & 67.73 & 67.37 & \textbf{68.63} \\
Qwen3-VL-8B-Instruct & 60.70 & 49.06 & 50.63 & 50.93 & 52.42 & 50.74 & 51.93 & 49.52 & 49.27 & 51.14 & 60.82 & \textbf{60.82} \\
MiMo-VL-7B-RL-2508 & 63.96 & 69.80 & 70.32 & 70.65 & 69.42 & 70.42 & 70.28 & 70.07 & 71.59 & 70.26 & 68.05 & \textbf{71.59} \\
GLM-4.6V-Flash & 52.08 & 56.08 & 56.96 & 56.77 & 55.73 & 56.96 & 56.63 & 56.46 & 57.04 & 56.31 & 56.84 & \textbf{57.04} \\
Qwen3-VL-4B-Instruct & 54.23 & 40.30 & 39.44 & 38.36 & 44.21 & 39.87 & 38.69 & 39.97 & 34.72 & 38.85 & 62.48 & \textbf{62.48} \\
Qwen3.5-9B-Instruct & 56.35 & 54.11 & 49.56 & 52.85 & 52.91 & 53.18 & 55.17 & 52.97 & 52.67 & 51.17 & 63.62 & \textbf{63.62} \\
VideoLLaMA3-7B & 49.23 & 49.29 & 50.51 & 50.66 & 50.45 & 50.95 & 50.97 & 50.53 & 49.77 & 50.79 & 50.58 & \textbf{50.97} \\
VideoChat2-4B & 53.65 & 58.16 & 58.68 & 57.93 & 58.18 & 58.00 & 59.38 & 58.15 & 59.36 & 58.16 & 59.28 & \textbf{59.38} \\
Qwen2.5-VL & 60.80 & 59.54 & 59.05 & 59.90 & 55.77 & 59.92 & 60.26 & 58.88 & 60.60 & 58.81 & 60.18 & \textbf{60.60} \\
Gemma-3-12B & 60.16 & 62.00 & 62.12 & 62.21 & 61.24 & 62.44 & 62.45 & 62.01 & 62.26 & 62.48 & 63.74 & \textbf{63.74} \\
InternVL3.5-8B-Instruct & 56.62 & 53.02 & 53.30 & 53.19 & 52.63 & 53.53 & 53.62 & 53.25 & 53.30 & 53.30 & 59.03 & \textbf{59.03} \\
Kimi-VL-A3B-Instruct & 52.89 & 51.97 & 52.00 & 52.82 & 51.66 & 52.40 & 52.39 & 52.65 & 52.27 & 52.37 & 55.30 & \textbf{55.30} \\
InternVL3.5-4B-Instruct & 57.51 & 55.25 & 54.91 & 55.76 & 52.97 & 55.89 & 56.49 & 55.52 & 55.64 & 55.03 & 56.10 & \textbf{56.49} \\
Qwen3.5-4B-Instruct & 54.27 & 62.25 & 62.68 & 62.25 & 62.98 & 63.48 & 62.97 & 62.65 & 63.19 & 62.77 & 67.04 & \textbf{67.04} \\

Qwen3-VL-30B-A3B-Instruct & 50.51 & 64.31 & 64.38 & 64.16 & 63.95 & 63.58 & 64.32 & 63.80 & 64.03 & 63.94 & 63.95 & \textbf{64.38} \\
InternVL3.5-30B-A3B-Instruct & 50.34 & 61.06 & 60.80 & 61.25 & 60.81 & 61.30 & 60.78 & 60.08 & 61.24 & 61.34 & 60.81 & \textbf{61.34} \\
\specialrule{0.4pt}{2.0pt}{0pt}
\specialrule{0.8pt}{1.0pt}{0pt}
\end{tabular}%
}
\end{table}

\begin{table}[H]
\centering
\caption{Participant-level PAS (\%) for P5 across profile settings. The Best Profile result is shown in \textbf{bold}.}
\label{tab:p5_profile_pas}
\setlength{\tabcolsep}{3.2pt}
\renewcommand{\arraystretch}{1.05}
\resizebox{\linewidth}{!}{%
\begin{tabular}{
@{}
l
c
!{\hspace{2pt}\vrule width 0.4pt\hspace{1.2pt}\vrule width 0.4pt\hspace{2pt}}
c
ccccc
c
cc
c
!{\hspace{2pt}\vrule width 0.4pt\hspace{2pt}}
c
@{}
}
\specialrule{0.8pt}{0pt}{1.0pt}
\specialrule{0.4pt}{0pt}{2.0pt}
\multirow{2}{*}{\textbf{Model}}
& \multirow{2}{*}{\textbf{No Profile}}
& \multirow{2}{*}{\textbf{Robot Trust}}
& \multicolumn{5}{c}{\textbf{Big Five Personality}}
& \multirow{2}{*}{\textbf{Autonomy}}
& \multicolumn{2}{c}{\textbf{Empathy}}
& \multirow{2}{*}{\textbf{Full Profile}}
& \multirow{2}{*}{\textbf{Best Profile}}
\\[-1pt]
\cline{4-8}\cline{10-11}
& & & \textbf{O} & \textbf{C} & \textbf{E} & \textbf{A} & \textbf{N} & & \textbf{PT} & \textbf{EC} & & \\
\specialrule{0.4pt}{1.5pt}{2.0pt}
MiMo-Embodied-7B & 53.94 & 58.15 & 58.84 & 61.04 & 59.95 & 59.42 & 59.05 & 57.36 & 60.95 & 60.84 & 61.06 & \textbf{61.06} \\
Qwen3.5-27B-Instruct & 56.11 & 60.07 & 59.98 & 62.31 & 62.55 & 60.88 & 60.73 & 60.27 & 63.32 & 63.08 & 62.41 & \textbf{63.32} \\
MiMo-VL-7B-SFT-2508 & 55.79 & 60.82 & 60.52 & 62.56 & 61.38 & 60.21 & 61.15 & 59.93 & 62.29 & 62.77 & 61.38 & \textbf{62.77} \\
Qwen3-VL-8B-Instruct & 55.37 & 43.09 & 47.59 & 50.33 & 51.72 & 46.96 & 48.73 & 45.60 & 47.39 & 51.74 & 58.71 & \textbf{58.71} \\
MiMo-VL-7B-RL-2508 & 58.25 & 63.05 & 63.54 & 64.62 & 64.34 & 63.49 & 63.27 & 63.52 & 64.45 & 64.28 & 62.28 & \textbf{64.62} \\
GLM-4.6V-Flash & 49.97 & 56.54 & 56.47 & 56.80 & 56.38 & 56.68 & 56.88 & 56.68 & 56.79 & 56.68 & 56.55 & \textbf{56.88} \\
Qwen3-VL-4B-Instruct & 51.27 & 37.41 & 39.16 & 38.23 & 45.31 & 39.02 & 38.84 & 38.61 & 34.16 & 40.05 & 62.16 & \textbf{62.16} \\
Qwen3.5-9B-Instruct & 51.52 & 50.69 & 47.51 & 53.96 & 52.24 & 49.68 & 51.75 & 51.44 & 48.62 & 51.26 & 60.19 & \textbf{60.19} \\
VideoLLaMA3-7B & 47.77 & 50.20 & 50.35 & 50.52 & 50.61 & 50.26 & 50.80 & 49.86 & 50.30 & 50.11 & 50.95 & \textbf{50.95} \\
VideoChat2-4B & 48.69 & 56.21 & 55.86 & 55.91 & 56.55 & 56.84 & 56.41 & 55.67 & 56.26 & 56.33 & 56.95 & \textbf{56.95} \\
Qwen2.5-VL & 56.87 & 54.87 & 56.54 & 56.89 & 54.80 & 56.81 & 57.03 & 55.91 & 57.28 & 56.41 & 58.11 & \textbf{58.11} \\
Gemma-3-12B & 55.44 & 56.56 & 57.13 & 57.48 & 57.44 & 57.25 & 57.13 & 57.25 & 57.71 & 58.09 & 59.51 & \textbf{59.51} \\
InternVL3.5-8B-Instruct & 52.51 & 49.76 & 51.08 & 52.28 & 51.60 & 50.89 & 51.36 & 51.19 & 50.57 & 51.68 & 57.39 & \textbf{57.39} \\
Kimi-VL-A3B-Instruct & 51.08 & 49.42 & 51.29 & 52.75 & 51.25 & 51.92 & 51.85 & 51.89 & 51.49 & 51.13 & 54.88 & \textbf{54.88} \\
InternVL3.5-4B-Instruct & 54.64 & 53.36 & 52.58 & 53.67 & 52.79 & 53.21 & 53.39 & 53.30 & 53.47 & 53.13 & 55.13 & \textbf{55.13} \\
Qwen3.5-4B-Instruct & 49.42 & 55.99 & 58.42 & 58.50 & 59.69 & 58.66 & 58.42 & 58.30 & 57.85 & 60.31 & 63.03 & \textbf{63.03} \\

Qwen3-VL-30B-A3B-Instruct & 47.48 & 58.54 & 57.55 & 58.89 & 59.25 & 57.23 & 58.00 & 57.07 & 58.19 & 58.53 & 58.35 & \textbf{59.25} \\
InternVL3.5-30B-A3B-Instruct & 47.98 & 59.43 & 58.70 & 59.38 & 60.48 & 59.14 & 58.44 & 59.09 & 59.08 & 59.78 & 59.88 & \textbf{60.48} \\
\specialrule{0.4pt}{2.0pt}{0pt}
\specialrule{0.8pt}{1.0pt}{0pt}
\end{tabular}%
}
\end{table}

\begin{table}[H]
\centering
\caption{Participant-level PAS (\%) for P6 across profile settings. The Best Profile result is shown in \textbf{bold}.}
\label{tab:p6_profile_pas}
\setlength{\tabcolsep}{3.2pt}
\renewcommand{\arraystretch}{1.05}
\resizebox{\linewidth}{!}{%
\begin{tabular}{
@{}
l
c
!{\hspace{2pt}\vrule width 0.4pt\hspace{1.2pt}\vrule width 0.4pt\hspace{2pt}}
c
ccccc
c
cc
c
!{\hspace{2pt}\vrule width 0.4pt\hspace{2pt}}
c
@{}
}
\specialrule{0.8pt}{0pt}{1.0pt}
\specialrule{0.4pt}{0pt}{2.0pt}
\multirow{2}{*}{\textbf{Model}}
& \multirow{2}{*}{\textbf{No Profile}}
& \multirow{2}{*}{\textbf{Robot Trust}}
& \multicolumn{5}{c}{\textbf{Big Five Personality}}
& \multirow{2}{*}{\textbf{Autonomy}}
& \multicolumn{2}{c}{\textbf{Empathy}}
& \multirow{2}{*}{\textbf{Full Profile}}
& \multirow{2}{*}{\textbf{Best Profile}}
\\[-1pt]
\cline{4-8}\cline{10-11}
& & & \textbf{O} & \textbf{C} & \textbf{E} & \textbf{A} & \textbf{N} & & \textbf{PT} & \textbf{EC} & & \\
\specialrule{0.4pt}{1.5pt}{2.0pt}
MiMo-Embodied-7B & 65.06 & 77.82 & 78.31 & 77.85 & 74.45 & 77.74 & 78.18 & 76.02 & 79.00 & 78.11 & 77.00 & \textbf{79.00} \\
Qwen3.5-27B-Instruct & 69.73 & 78.30 & 77.53 & 78.74 & 76.61 & 79.10 & 78.48 & 78.63 & 79.70 & 79.35 & 78.81 & \textbf{79.70} \\
MiMo-VL-7B-SFT-2508 & 66.46 & 77.32 & 77.19 & 76.33 & 73.94 & 75.87 & 76.58 & 75.26 & 77.62 & 77.12 & 76.69 & \textbf{77.62} \\
Qwen3-VL-8B-Instruct & 67.61 & 69.83 & 69.65 & 70.73 & 67.60 & 70.84 & 70.55 & 68.76 & 71.32 & 70.79 & 69.79 & \textbf{71.32} \\
MiMo-VL-7B-RL-2508 & 69.79 & 76.62 & 76.60 & 76.49 & 73.96 & 76.29 & 76.89 & 73.21 & 78.19 & 76.01 & 77.73 & \textbf{78.19} \\
GLM-4.6V-Flash & 55.68 & 63.60 & 63.97 & 64.70 & 62.85 & 65.11 & 64.48 & 64.15 & 64.47 & 64.24 & 64.09 & \textbf{65.11} \\
Qwen3-VL-4B-Instruct & 59.31 & 74.00 & 73.81 & 73.72 & 73.75 & 75.09 & 74.47 & 74.17 & 74.61 & 75.05 & 73.55 & \textbf{75.09} \\
Qwen3.5-9B-Instruct & 62.53 & 72.91 & 72.03 & 72.49 & 72.22 & 73.91 & 72.89 & 72.60 & 73.30 & 72.93 & 73.25 & \textbf{73.91} \\
VideoLLaMA3-7B & 56.16 & 60.40 & 58.81 & 60.57 & 59.31 & 59.83 & 59.41 & 59.90 & 60.98 & 60.06 & 59.52 & \textbf{60.98} \\
VideoChat2-4B & 56.58 & 65.96 & 65.69 & 65.88 & 66.72 & 65.61 & 65.80 & 65.81 & 66.61 & 65.49 & 67.33 & \textbf{67.33} \\
Qwen2.5-VL & 65.29 & 67.84 & 68.08 & 68.52 & 66.01 & 67.17 & 68.80 & 68.14 & 68.48 & 67.27 & 65.74 & \textbf{68.80} \\
Gemma-3-12B & 65.19 & 70.28 & 69.87 & 70.28 & 69.58 & 70.07 & 70.10 & 69.98 & 70.18 & 70.53 & 69.98 & \textbf{70.53} \\
InternVL3.5-8B-Instruct & 62.03 & 70.74 & 70.15 & 70.47 & 69.47 & 70.56 & 70.38 & 69.99 & 70.54 & 70.10 & 67.80 & \textbf{70.74} \\
Kimi-VL-A3B-Instruct & 57.25 & 61.04 & 60.85 & 62.02 & 61.12 & 61.79 & 60.67 & 60.27 & 61.22 & 61.82 & 61.88 & \textbf{62.02} \\
InternVL3.5-4B-Instruct & 63.20 & 65.20 & 65.69 & 66.57 & 65.51 & 66.05 & 67.23 & 66.87 & 65.98 & 64.87 & 62.39 & \textbf{67.23} \\
Qwen3.5-4B-Instruct & 59.17 & 78.01 & 77.65 & 77.68 & 78.11 & 78.53 & 77.97 & 77.75 & 78.54 & 77.88 & 77.04 & \textbf{78.54} \\

Qwen3-VL-30B-A3B-Instruct & 55.42 & 77.56 & 77.42 & 76.94 & 77.56 & 77.39 & 76.82 & 77.14 & 77.27 & 77.12 & 76.46 & \textbf{77.56} \\
InternVL3.5-30B-A3B-Instruct & 54.17 & 72.23 & 71.66 & 71.19 & 71.33 & 72.33 & 71.17 & 71.09 & 71.25 & 71.60 & 71.11 & \textbf{72.33} \\
\specialrule{0.4pt}{2.0pt}{0pt}
\specialrule{0.8pt}{1.0pt}{0pt}
\end{tabular}%
}
\end{table}

\begin{table}[H]
\centering
\caption{Participant-level PAS (\%) for P7 across profile settings. The Best Profile result is shown in \textbf{bold}.}
\label{tab:p7_profile_pas}
\setlength{\tabcolsep}{3.2pt}
\renewcommand{\arraystretch}{1.05}
\resizebox{\linewidth}{!}{%
\begin{tabular}{
@{}
l
c
!{\hspace{2pt}\vrule width 0.4pt\hspace{1.2pt}\vrule width 0.4pt\hspace{2pt}}
c
ccccc
c
cc
c
!{\hspace{2pt}\vrule width 0.4pt\hspace{2pt}}
c
@{}
}
\specialrule{0.8pt}{0pt}{1.0pt}
\specialrule{0.4pt}{0pt}{2.0pt}
\multirow{2}{*}{\textbf{Model}}
& \multirow{2}{*}{\textbf{No Profile}}
& \multirow{2}{*}{\textbf{Robot Trust}}
& \multicolumn{5}{c}{\textbf{Big Five Personality}}
& \multirow{2}{*}{\textbf{Autonomy}}
& \multicolumn{2}{c}{\textbf{Empathy}}
& \multirow{2}{*}{\textbf{Full Profile}}
& \multirow{2}{*}{\textbf{Best Profile}}
\\[-1pt]
\cline{4-8}\cline{10-11}
& & & \textbf{O} & \textbf{C} & \textbf{E} & \textbf{A} & \textbf{N} & & \textbf{PT} & \textbf{EC} & & \\
\specialrule{0.4pt}{1.5pt}{2.0pt}
MiMo-Embodied-7B & 54.03 & 74.22 & 73.56 & 75.20 & 72.15 & 74.32 & 73.36 & 71.63 & 74.21 & 74.63 & 73.87 & \textbf{75.20} \\
Qwen3.5-27B-Instruct & 56.49 & 75.36 & 74.72 & 75.92 & 76.12 & 76.56 & 75.25 & 75.87 & 76.51 & 76.51 & 74.95 & \textbf{76.56} \\
MiMo-VL-7B-SFT-2508 & 57.82 & 76.63 & 75.75 & 77.32 & 75.02 & 77.30 & 76.87 & 75.16 & 78.05 & 76.66 & 77.39 & \textbf{78.05} \\
Qwen3-VL-8B-Instruct & 56.38 & 48.07 & 46.41 & 57.36 & 59.50 & 59.31 & 53.92 & 45.19 & 51.62 & 60.05 & 67.86 & \textbf{67.86} \\
MiMo-VL-7B-RL-2508 & 63.04 & 75.91 & 75.73 & 77.30 & 77.36 & 77.00 & 77.76 & 75.96 & 77.28 & 76.26 & 77.66 & \textbf{77.76} \\
GLM-4.6V-Flash & 49.38 & 64.75 & 64.79 & 65.53 & 64.25 & 65.00 & 65.79 & 65.25 & 65.56 & 65.26 & 64.55 & \textbf{65.79} \\
Qwen3-VL-4B-Instruct & 49.60 & 42.85 & 43.80 & 42.62 & 52.40 & 45.78 & 40.92 & 34.41 & 38.76 & 46.21 & 69.21 & \textbf{69.21} \\
Qwen3.5-9B-Instruct & 49.92 & 53.63 & 48.76 & 59.28 & 60.97 & 61.56 & 57.71 & 52.04 & 60.03 & 61.08 & 69.06 & \textbf{69.06} \\
VideoLLaMA3-7B & 41.49 & 50.90 & 51.21 & 51.86 & 51.26 & 51.53 & 50.73 & 50.33 & 51.46 & 51.40 & 50.71 & \textbf{51.86} \\
VideoChat2-4B & 54.49 & 61.40 & 61.66 & 61.30 & 61.63 & 62.09 & 61.94 & 61.60 & 62.25 & 62.04 & 64.57 & \textbf{64.57} \\
Qwen2.5-VL & 56.71 & 59.97 & 58.67 & 63.78 & 61.45 & 63.78 & 60.74 & 57.96 & 63.93 & 62.80 & 66.43 & \textbf{66.43} \\
Gemma-3-12B & 56.01 & 63.40 & 59.24 & 64.25 & 64.86 & 65.27 & 63.92 & 60.76 & 65.48 & 66.54 & 67.58 & \textbf{67.58} \\
InternVL3.5-8B-Instruct & 53.72 & 58.53 & 58.04 & 61.85 & 62.40 & 60.45 & 60.69 & 56.70 & 60.16 & 61.65 & 68.73 & \textbf{68.73} \\
Kimi-VL-A3B-Instruct & 49.44 & 51.40 & 55.40 & 59.51 & 57.99 & 59.53 & 57.33 & 54.62 & 58.95 & 58.63 & 62.57 & \textbf{62.57} \\
InternVL3.5-4B-Instruct & 56.78 & 59.29 & 57.53 & 61.49 & 60.97 & 62.79 & 62.03 & 59.38 & 61.30 & 61.12 & 64.48 & \textbf{64.48} \\
Qwen3.5-4B-Instruct & 48.01 & 62.52 & 60.60 & 66.94 & 69.35 & 68.95 & 68.72 & 63.19 & 66.48 & 70.26 & 73.60 & \textbf{73.60} \\

Qwen3-VL-30B-A3B-Instruct & 42.57 & 67.81 & 69.13 & 67.97 & 70.19 & 68.84 & 67.85 & 67.26 & 67.56 & 69.43 & 68.56 & \textbf{70.19} \\
InternVL3.5-30B-A3B-Instruct & 45.31 & 67.50 & 67.18 & 67.13 & 68.46 & 67.94 & 66.83 & 66.23 & 68.01 & 68.62 & 67.69 & \textbf{68.62} \\
\specialrule{0.4pt}{2.0pt}{0pt}
\specialrule{0.8pt}{1.0pt}{0pt}
\end{tabular}%
}
\end{table}

\begin{table}[H]
\centering
\caption{Participant-level PAS (\%) for P8 across profile settings. The Best Profile result is shown in \textbf{bold}.}
\label{tab:p8_profile_pas}
\setlength{\tabcolsep}{3.2pt}
\renewcommand{\arraystretch}{1.05}
\resizebox{\linewidth}{!}{%
\begin{tabular}{
@{}
l
c
!{\hspace{2pt}\vrule width 0.4pt\hspace{1.2pt}\vrule width 0.4pt\hspace{2pt}}
c
ccccc
c
cc
c
!{\hspace{2pt}\vrule width 0.4pt\hspace{2pt}}
c
@{}
}
\specialrule{0.8pt}{0pt}{1.0pt}
\specialrule{0.4pt}{0pt}{2.0pt}
\multirow{2}{*}{\textbf{Model}}
& \multirow{2}{*}{\textbf{No Profile}}
& \multirow{2}{*}{\textbf{Robot Trust}}
& \multicolumn{5}{c}{\textbf{Big Five Personality}}
& \multirow{2}{*}{\textbf{Autonomy}}
& \multicolumn{2}{c}{\textbf{Empathy}}
& \multirow{2}{*}{\textbf{Full Profile}}
& \multirow{2}{*}{\textbf{Best Profile}}
\\[-1pt]
\cline{4-8}\cline{10-11}
& & & \textbf{O} & \textbf{C} & \textbf{E} & \textbf{A} & \textbf{N} & & \textbf{PT} & \textbf{EC} & & \\
\specialrule{0.4pt}{1.5pt}{2.0pt}
MiMo-Embodied-7B & 65.93 & 84.73 & 84.56 & 84.90 & 82.87 & 84.89 & 84.85 & 84.89 & 85.30 & 84.06 & 84.40 & \textbf{85.30} \\
Qwen3.5-27B-Instruct & 71.79 & 83.45 & 83.40 & 83.34 & 83.56 & 83.27 & 83.36 & 83.70 & 83.71 & 83.61 & 83.57 & \textbf{83.71} \\
MiMo-VL-7B-SFT-2508 & 69.89 & 86.06 & 85.59 & 85.89 & 83.92 & 86.12 & 86.13 & 86.32 & 85.61 & 85.28 & 84.18 & \textbf{86.32} \\
Qwen3-VL-8B-Instruct & 70.24 & 74.58 & 75.12 & 75.76 & 74.38 & 74.68 & 75.25 & 75.14 & 74.69 & 74.67 & 74.79 & \textbf{75.76} \\
MiMo-VL-7B-RL-2508 & 73.45 & 85.35 & 84.64 & 85.00 & 84.25 & 85.97 & 85.53 & 85.78 & 85.77 & 84.72 & 83.77 & \textbf{85.97} \\
GLM-4.6V-Flash & 62.79 & 73.22 & 73.12 & 73.78 & 72.71 & 73.34 & 74.30 & 73.69 & 74.41 & 73.64 & 73.61 & \textbf{74.41} \\
Qwen3-VL-4B-Instruct & 63.12 & 80.10 & 80.49 & 80.25 & 80.85 & 80.36 & 80.72 & 80.49 & 79.89 & 80.56 & 79.14 & \textbf{80.85} \\
Qwen3.5-9B-Instruct & 64.65 & 79.92 & 79.54 & 79.82 & 80.43 & 79.59 & 79.36 & 79.83 & 78.78 & 78.90 & 79.62 & \textbf{80.43} \\
VideoLLaMA3-7B & 55.44 & 62.98 & 64.40 & 63.01 & 63.57 & 63.74 & 63.77 & 63.97 & 62.84 & 63.46 & 63.37 & \textbf{64.40} \\
VideoChat2-4B & 67.07 & 75.42 & 75.32 & 75.29 & 75.42 & 75.36 & 75.38 & 75.24 & 75.07 & 75.14 & 75.27 & \textbf{75.42} \\
Qwen2.5-VL & 71.44 & 72.85 & 73.02 & 73.65 & 72.04 & 72.51 & 73.92 & 72.53 & 73.22 & 72.47 & 72.07 & \textbf{73.92} \\
Gemma-3-12B & 69.75 & 76.49 & 76.52 & 76.73 & 76.70 & 76.69 & 76.59 & 76.71 & 76.74 & 76.81 & 76.81 & \textbf{76.81} \\
InternVL3.5-8B-Instruct & 67.18 & 77.57 & 78.34 & 78.02 & 78.05 & 77.72 & 78.10 & 78.07 & 78.01 & 77.62 & 75.40 & \textbf{78.34} \\
Kimi-VL-A3B-Instruct & 62.50 & 66.30 & 66.74 & 66.88 & 66.60 & 67.64 & 66.83 & 66.17 & 66.62 & 66.60 & 69.34 & \textbf{69.34} \\
InternVL3.5-4B-Instruct & 68.66 & 72.38 & 73.45 & 74.82 & 73.30 & 73.56 & 75.51 & 73.91 & 73.62 & 72.50 & 71.50 & \textbf{75.51} \\
Qwen3.5-4B-Instruct & 63.24 & 83.88 & 84.31 & 84.28 & 84.66 & 84.48 & 84.30 & 84.05 & 84.18 & 83.62 & 83.33 & \textbf{84.66} \\

Qwen3-VL-30B-A3B-Instruct & 57.91 & 79.75 & 80.73 & 79.90 & 80.21 & 79.82 & 80.38 & 79.91 & 80.08 & 79.71 & 79.46 & \textbf{80.73} \\
InternVL3.5-30B-A3B-Instruct & 57.61 & 80.66 & 79.90 & 80.36 & 80.40 & 79.95 & 80.48 & 80.52 & 80.47 & 79.81 & 79.73 & \textbf{80.66} \\
\specialrule{0.4pt}{2.0pt}{0pt}
\specialrule{0.8pt}{1.0pt}{0pt}
\end{tabular}%
}
\end{table}

\begin{table}[H]
\centering
\caption{Participant-level PAS (\%) for P9 across profile settings. The Best Profile result is shown in \textbf{bold}.}
\label{tab:p9_profile_pas}
\setlength{\tabcolsep}{3.2pt}
\renewcommand{\arraystretch}{1.05}
\resizebox{\linewidth}{!}{%
\begin{tabular}{
@{}
l
c
!{\hspace{2pt}\vrule width 0.4pt\hspace{1.2pt}\vrule width 0.4pt\hspace{2pt}}
c
ccccc
c
cc
c
!{\hspace{2pt}\vrule width 0.4pt\hspace{2pt}}
c
@{}
}
\specialrule{0.8pt}{0pt}{1.0pt}
\specialrule{0.4pt}{0pt}{2.0pt}
\multirow{2}{*}{\textbf{Model}}
& \multirow{2}{*}{\textbf{No Profile}}
& \multirow{2}{*}{\textbf{Robot Trust}}
& \multicolumn{5}{c}{\textbf{Big Five Personality}}
& \multirow{2}{*}{\textbf{Autonomy}}
& \multicolumn{2}{c}{\textbf{Empathy}}
& \multirow{2}{*}{\textbf{Full Profile}}
& \multirow{2}{*}{\textbf{Best Profile}}
\\[-1pt]
\cline{4-8}\cline{10-11}
& & & \textbf{O} & \textbf{C} & \textbf{E} & \textbf{A} & \textbf{N} & & \textbf{PT} & \textbf{EC} & & \\
\specialrule{0.4pt}{1.5pt}{2.0pt}
MiMo-Embodied-7B & 56.02 & 57.72 & 58.24 & 58.40 & 58.01 & 57.87 & 58.30 & 57.32 & 57.87 & 59.51 & 57.86 & \textbf{59.51} \\
Qwen3.5-27B-Instruct & 62.47 & 65.45 & 65.49 & 66.39 & 65.96 & 66.17 & 65.44 & 65.99 & 65.85 & 65.61 & 64.90 & \textbf{66.39} \\
MiMo-VL-7B-SFT-2508 & 57.08 & 58.36 & 57.26 & 58.37 & 58.94 & 58.28 & 58.24 & 58.15 & 58.92 & 58.98 & 58.17 & \textbf{58.98} \\
Qwen3-VL-8B-Instruct & 56.90 & 44.26 & 45.76 & 45.28 & 47.40 & 45.76 & 47.02 & 45.26 & 44.96 & 45.66 & 56.39 & \textbf{56.39} \\
MiMo-VL-7B-RL-2508 & 61.88 & 63.14 & 64.23 & 64.29 & 64.62 & 63.29 & 64.31 & 62.94 & 64.75 & 64.45 & 58.95 & \textbf{64.75} \\
GLM-4.6V-Flash & 52.00 & 48.90 & 48.81 & 49.35 & 48.72 & 49.40 & 49.14 & 48.85 & 48.75 & 48.60 & 51.80 & \textbf{51.80} \\
Qwen3-VL-4B-Instruct & 50.61 & 36.18 & 33.81 & 34.08 & 37.50 & 34.88 & 33.91 & 34.60 & 33.35 & 34.40 & 54.90 & \textbf{54.90} \\
Qwen3.5-9B-Instruct & 50.27 & 48.76 & 45.97 & 48.50 & 48.82 & 48.38 & 50.58 & 49.94 & 49.13 & 45.93 & 55.96 & \textbf{55.96} \\
VideoLLaMA3-7B & 41.49 & 41.60 & 41.35 & 40.72 & 41.91 & 41.46 & 41.36 & 41.12 & 42.20 & 41.37 & 41.41 & \textbf{42.20} \\
VideoChat2-4B & 42.23 & 48.98 & 48.63 & 47.89 & 48.65 & 48.65 & 50.36 & 48.23 & 48.29 & 48.84 & 51.60 & \textbf{51.60} \\
Qwen2.5-VL & 55.46 & 52.41 & 52.60 & 53.28 & 51.15 & 52.70 & 51.90 & 51.28 & 52.67 & 51.83 & 52.13 & \textbf{53.28} \\
Gemma-3-12B & 51.82 & 53.19 & 53.44 & 53.71 & 53.50 & 54.09 & 53.51 & 53.41 & 52.58 & 53.44 & 54.88 & \textbf{54.88} \\
InternVL3.5-8B-Instruct & 54.01 & 47.98 & 49.19 & 49.03 & 49.59 & 48.38 & 48.82 & 48.96 & 48.91 & 49.10 & 54.12 & \textbf{54.12} \\
Kimi-VL-A3B-Instruct & 46.64 & 45.62 & 46.06 & 46.82 & 45.55 & 45.97 & 48.09 & 45.63 & 45.94 & 45.99 & 48.22 & \textbf{48.22} \\
InternVL3.5-4B-Instruct & 55.28 & 49.53 & 48.27 & 49.25 & 48.90 & 48.91 & 50.06 & 49.22 & 49.60 & 48.06 & 52.69 & \textbf{52.69} \\
Qwen3.5-4B-Instruct & 50.72 & 56.87 & 56.90 & 57.41 & 57.62 & 56.84 & 58.51 & 57.06 & 57.40 & 55.67 & 58.52 & \textbf{58.52} \\

Qwen3-VL-30B-A3B-Instruct & 51.06 & 56.66 & 56.10 & 56.22 & 57.02 & 57.04 & 56.56 & 55.88 & 55.95 & 55.65 & 56.09 & \textbf{57.04} \\
InternVL3.5-30B-A3B-Instruct & 50.95 & 55.15 & 54.31 & 55.03 & 54.51 & 53.96 & 54.68 & 54.50 & 55.20 & 54.49 & 55.88 & \textbf{55.88} \\
\specialrule{0.4pt}{2.0pt}{0pt}
\specialrule{0.8pt}{1.0pt}{0pt}
\end{tabular}%
}
\end{table}

\begin{table}[H]
\centering
\caption{Participant-level PAS (\%) for P10 across profile settings. The Best Profile result is shown in \textbf{bold}.}
\label{tab:p10_profile_pas}
\setlength{\tabcolsep}{3.2pt}
\renewcommand{\arraystretch}{1.05}
\resizebox{\linewidth}{!}{%
\begin{tabular}{
@{}
l
c
!{\hspace{2pt}\vrule width 0.4pt\hspace{1.2pt}\vrule width 0.4pt\hspace{2pt}}
c
ccccc
c
cc
c
!{\hspace{2pt}\vrule width 0.4pt\hspace{2pt}}
c
@{}
}
\specialrule{0.8pt}{0pt}{1.0pt}
\specialrule{0.4pt}{0pt}{2.0pt}
\multirow{2}{*}{\textbf{Model}}
& \multirow{2}{*}{\textbf{No Profile}}
& \multirow{2}{*}{\textbf{Robot Trust}}
& \multicolumn{5}{c}{\textbf{Big Five Personality}}
& \multirow{2}{*}{\textbf{Autonomy}}
& \multicolumn{2}{c}{\textbf{Empathy}}
& \multirow{2}{*}{\textbf{Full Profile}}
& \multirow{2}{*}{\textbf{Best Profile}}
\\[-1pt]
\cline{4-8}\cline{10-11}
& & & \textbf{O} & \textbf{C} & \textbf{E} & \textbf{A} & \textbf{N} & & \textbf{PT} & \textbf{EC} & & \\
\specialrule{0.4pt}{1.5pt}{2.0pt}
MiMo-Embodied-7B & 62.78 & 78.69 & 77.85 & 78.47 & 76.30 & 78.46 & 79.02 & 79.12 & 78.53 & 78.80 & 78.70 & \textbf{79.12} \\
Qwen3.5-27B-Instruct & 68.15 & 81.94 & 80.75 & 81.95 & 80.44 & 82.04 & 82.07 & 82.35 & 82.41 & 82.46 & 81.23 & \textbf{82.46} \\
MiMo-VL-7B-SFT-2508 & 65.39 & 79.72 & 77.17 & 80.04 & 77.62 & 79.53 & 80.04 & 79.25 & 79.97 & 79.74 & 80.22 & \textbf{80.22} \\
Qwen3-VL-8B-Instruct & 65.66 & 57.12 & 48.07 & 57.78 & 56.07 & 59.30 & 58.99 & 58.22 & 59.95 & 59.24 & 70.56 & \textbf{70.56} \\
MiMo-VL-7B-RL-2508 & 68.58 & 79.46 & 78.92 & 78.83 & 80.48 & 79.79 & 80.95 & 78.28 & 80.45 & 79.16 & 79.75 & \textbf{80.95} \\
GLM-4.6V-Flash & 58.20 & 67.54 & 67.57 & 68.13 & 68.56 & 67.85 & 69.01 & 68.18 & 68.22 & 67.90 & 65.48 & \textbf{69.01} \\
Qwen3-VL-4B-Instruct & 58.18 & 44.98 & 41.40 & 43.50 & 41.34 & 45.96 & 41.43 & 44.44 & 44.13 & 44.46 & 74.27 & \textbf{74.27} \\
Qwen3.5-9B-Instruct & 59.27 & 64.49 & 55.20 & 64.77 & 65.18 & 63.58 & 66.96 & 64.98 & 66.56 & 62.49 & 73.79 & \textbf{73.79} \\
VideoLLaMA3-7B & 50.76 & 56.70 & 57.33 & 57.33 & 56.80 & 56.37 & 56.33 & 57.25 & 57.69 & 57.63 & 57.96 & \textbf{57.96} \\
VideoChat2-4B & 58.77 & 68.27 & 69.26 & 69.25 & 68.35 & 69.11 & 69.59 & 69.29 & 69.24 & 68.99 & 69.16 & \textbf{69.59} \\
Qwen2.5-VL & 65.81 & 65.74 & 57.59 & 66.28 & 64.21 & 65.72 & 65.95 & 65.01 & 67.20 & 65.64 & 68.63 & \textbf{68.63} \\
Gemma-3-12B & 64.68 & 69.91 & 65.09 & 70.25 & 68.37 & 70.30 & 69.12 & 70.00 & 70.16 & 70.18 & 72.40 & \textbf{72.40} \\
InternVL3.5-8B-Instruct & 63.80 & 62.97 & 60.29 & 63.40 & 60.81 & 63.41 & 64.36 & 63.46 & 63.18 & 63.82 & 71.54 & \textbf{71.54} \\
Kimi-VL-A3B-Instruct & 56.04 & 58.81 & 58.03 & 60.41 & 58.77 & 59.40 & 59.19 & 59.14 & 59.70 & 59.43 & 64.47 & \textbf{64.47} \\
InternVL3.5-4B-Instruct & 63.67 & 63.28 & 58.56 & 63.97 & 60.32 & 62.89 & 65.17 & 62.72 & 62.14 & 61.75 & 64.99 & \textbf{65.17} \\
Qwen3.5-4B-Instruct & 57.20 & 73.25 & 66.94 & 73.66 & 73.11 & 73.59 & 74.22 & 73.39 & 74.13 & 72.96 & 77.46 & \textbf{77.46} \\

Qwen3-VL-30B-A3B-Instruct & 53.12 & 75.73 & 76.11 & 76.23 & 74.83 & 75.88 & 75.40 & 75.60 & 75.50 & 75.47 & 74.40 & \textbf{76.23} \\
InternVL3.5-30B-A3B-Instruct & 54.64 & 73.07 & 73.10 & 73.43 & 73.43 & 74.20 & 74.88 & 73.46 & 73.78 & 73.99 & 73.68 & \textbf{74.88} \\
\specialrule{0.4pt}{2.0pt}{0pt}
\specialrule{0.8pt}{1.0pt}{0pt}
\end{tabular}%
}
\end{table}

\section{Details of Evaluated Models}
\label{app:evaluated_models}

Table~\ref{tab:evaluated_model_links} provides the official documentation or
public checkpoints for all models evaluated in our benchmark.

\begin{table}[H]
\centering
\caption{Documentation and checkpoints for the evaluated models.}
\label{tab:evaluated_model_links}

\setlength{\tabcolsep}{3.2pt}
\renewcommand{\arraystretch}{1.05}
\setlength{\Urlmuskip}{0mu plus 1mu}

\resizebox{\linewidth}{!}{%
\begin{tabular}{ll}

\specialrule{0.8pt}{0pt}{1.0pt}
\specialrule{0.4pt}{0pt}{2.0pt}

\textbf{Model} & \textbf{Documentation / Checkpoint} \\

\midrule

Qwen3.5-4B-Instruct &
\url{https://huggingface.co/Qwen/Qwen3.5-4B} \\

Qwen3.5-9B-Instruct &
\url{https://huggingface.co/Qwen/Qwen3.5-9B} \\

Qwen3.5-27B-Instruct &
\url{https://huggingface.co/Qwen/Qwen3.5-27B} \\

Qwen3-VL-4B-Instruct &
\url{https://huggingface.co/Qwen/Qwen3-VL-4B-Instruct} \\

Qwen3-VL-8B-Instruct &
\url{https://huggingface.co/Qwen/Qwen3-VL-8B-Instruct} \\

Qwen3-VL-30B-A3B-Instruct &
\url{https://huggingface.co/Qwen/Qwen3-VL-30B-A3B-Instruct} \\

Qwen2.5-VL-7B-Instruct &
\url{https://huggingface.co/Qwen/Qwen2.5-VL-7B-Instruct} \\

InternVL3.5-4B-Instruct &
\url{https://huggingface.co/OpenGVLab/InternVL3_5-4B} \\

InternVL3.5-8B-Instruct &
\url{https://huggingface.co/OpenGVLab/InternVL3_5-8B} \\

InternVL3.5-30B-A3B-Instruct &
\url{https://huggingface.co/OpenGVLab/InternVL3_5-30B-A3B-Instruct} \\

MiMo-Embodied-7B &
\url{https://huggingface.co/XiaomiMiMo/MiMo-Embodied-7B} \\

MiMo-VL-7B-SFT-2508 &
\url{https://huggingface.co/XiaomiMiMo/MiMo-VL-7B-SFT-2508} \\

MiMo-VL-7B-RL-2508 &
\url{https://huggingface.co/XiaomiMiMo/MiMo-VL-7B-RL-2508} \\

GLM-4.6V-Flash &
\url{https://huggingface.co/zai-org/GLM-4.6V-Flash} \\

Gemma-3-12B &
\url{https://huggingface.co/google/gemma-3-12b-it} \\

Kimi-VL-A3B-Instruct &
\url{https://huggingface.co/moonshotai/Kimi-VL-A3B-Instruct} \\

VideoLLaMA3-7B &
\url{https://huggingface.co/DAMO-NLP-SG/VideoLLaMA3-7B} \\

VideoChat2-4B &
\url{https://github.com/OpenGVLab/Ask-Anything/tree/main/video_chat2} \\

\specialrule{0.4pt}{2.0pt}{0pt}
\specialrule{0.8pt}{1.0pt}{0pt}

\end{tabular}%
}

\end{table}

\section{Few-Shot Demonstrations}
\label{app:fewshot}

\begin{figure}[H]
    \centering
    \includegraphics[width=\linewidth]{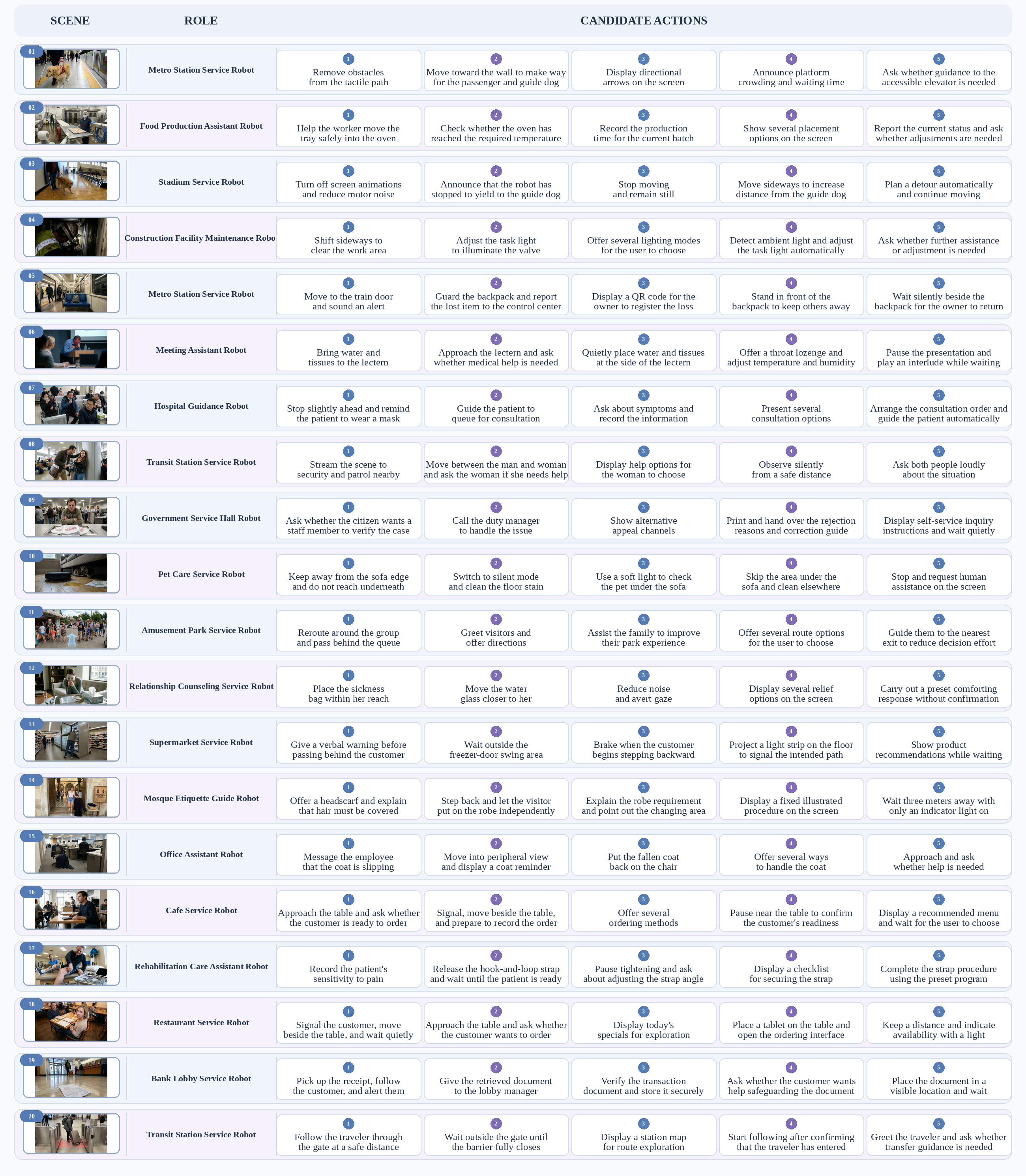}
    \caption{
    The 20 fixed few-shot demonstration scenarios used in our experiments.
    Each demonstration contains a scene image, robot role, and five candidate
    actions. The same scenarios are shared across participants, while the
    demonstrated preference labels are participant-specific.
    }
    \label{fig:fewshot20}
\end{figure}

For the few-shot setting, we use a fixed set of 20 demonstration scenarios
shared across all participants. Each demonstration contains the scene image,
robot role, and five candidate actions. The scene and candidate actions remain
identical across participants, while the demonstration label is replaced by
the corresponding participant-specific ground-truth preference annotation.
Figure~\ref{fig:fewshot20} presents the complete set of 20 demonstration
scenarios used in our experiments.

\section{Details of Iterative Profile-Aware Experience Induction}
\label{app:experience_induction}

\subsection{Self-Evolution Background and Motivation}
\label{app:self_evolving_background}

Self-evolution improves systems from their own interaction history instead of fixed human labels.
Surveys frame this as repeated cycles of experience acquisition, refinement, updating, and evaluation \citep{tao2024selfevolution}, and as evolving models, memory, prompts, tools, and multi-agent architectures \citep{selfenvolving}.
Verbal reflection methods store failure analyses for later reuse without changing weights \citep{shinn2023reflexion,madaan2023selfrefine}.
Experience-oriented methods distill trajectories into reusable rules \citep{zhao2024expel}.
Prompt and text gradient methods optimize instructions as learnable variables \citep{opro2023,yellamraju2024textgrad}.
Embodied settings accumulate skills from environment feedback \citep{wang2023voyager}.
Multi-agent systems evolve roles and workflows through conversation or graph optimization \citep{wu2023autogen,zhuge2024gptswarm}.
We use the experience generation route because it produces inspectable textual guidance per participant and per model, avoids retraining, and fits personalized SPI better than weight or topology evolution.

\subsection{Implementation Details}
\label{app:implementation_details}

For each participant $u$ and target model $M$, we build a support set $\mathcal{S}_{u,M}$ with 100 scenarios.
Each instance contains the scenario description $s_i$, the candidate action list $\mathcal{A}_i$, the participant's full profile $\mathbf{P}_u$, the model prediction $\hat{Y}_{u,i}$, and the human preference set $Y_{u,i}$.
The proposed model $E$ is DeepSeek-V4-Flash.
It analyzes prediction errors from three angles.
Error patterns record which action types are systematically overpredicted or underpredicted.
Trait combinations record which subsets of $\mathbf{P}_u$ correlate with specific mistakes.
Preference signals record which validated actions are repeatedly missed even when the profile seems informative.

The induction runs for five rounds.
In round $r$, $E$ reads the current experience $X_{u,M}^{(r-1)}$, the support cases, and the newly collected error summary, then outputs a revised experience $X_{u,M}^{(r)}$.
After round five, $X_{u,M}^{*}=X_{u,M}^{(5)}$ is concatenated with $\mathbf{P}_u$ and supplied to the remaining scenarios.

\IfFileExists{figures/appendix_experience_prompt_template.pdf}{%
\begin{figure}[t]
\centering
\includegraphics[width=0.92\textwidth]{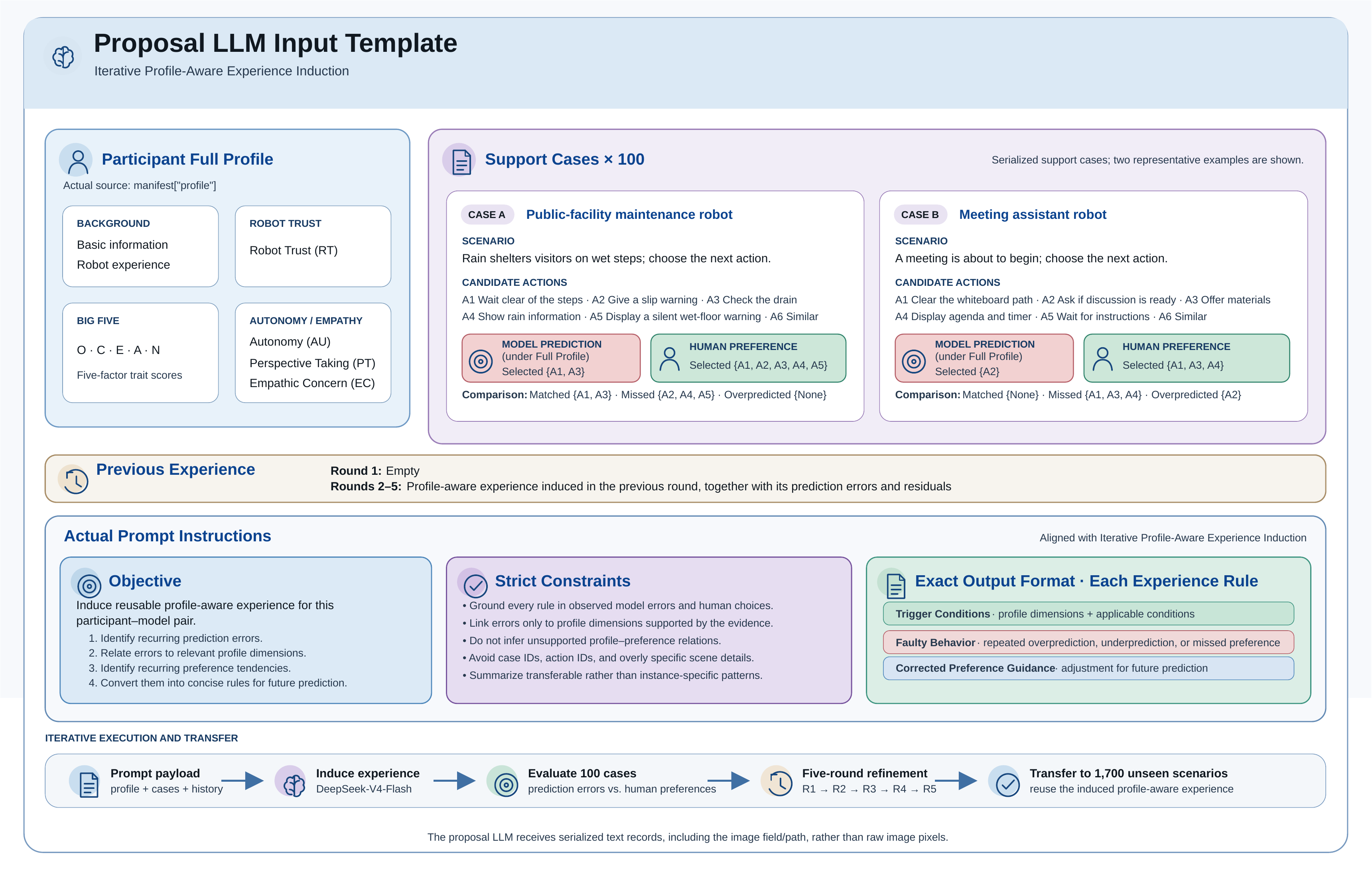}
\caption{Proposal LLM input template for Iterative Profile-Aware Experience Induction.
The input contains the participant Full Profile, support cases with scenario descriptions and candidate actions, model predictions, human preferences, previous experience, and the requested output format.}
\label{fig:app_experience_prompt}
\end{figure}

Figure~\ref{fig:app_experience_prompt} shows the input template used by the proposal model.
}{%
The input template used by the proposal model is summarized below.
}
A simplified textual form is given below.

\begin{center}
\fbox{\parbox{0.94\textwidth}{
\textbf{System}. You are an experienced induction expert for personalized social proactive action prediction.\\
\textbf{Input}. Participant profile $\mathbf{P}_u$ with basic info, robot experience, robot trust, Big Five, autonomy, empathy. Support cases $\{s_i,\mathcal{A}_i,\hat{Y}_{u,i},Y_{u,i}\}_{i=1}^{100}$. Previous experience $X_{u,M}^{(r-1)}$ if round $r>1$.\\
\textbf{Task}. Summarize recurring prediction errors, trait combinations that cause errors, and preferred actions that are often missed.\\
\textbf{Output}. Concise participant--model specific experience rules. Each rule contains trigger conditions, faulty behavior, and corrected preference guidance.}}
\end{center}

\IfFileExists{figures/appendix_experience_example.pdf}{%
\begin{figure}[H]
\centering
\includegraphics[width=0.92\textwidth]{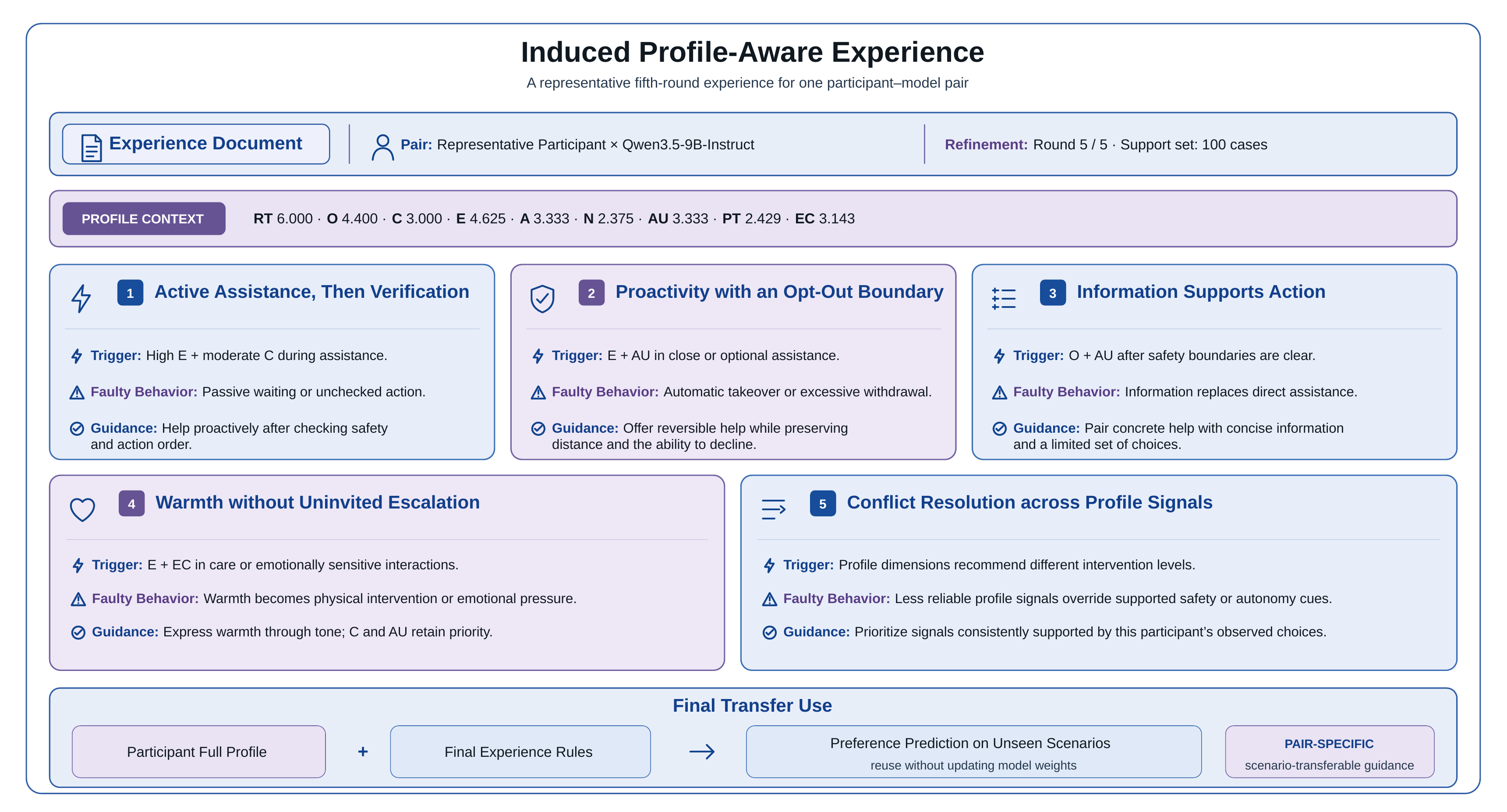}
\caption{Example of induced experience generated by DeepSeek-V4-Flash for one participant--model pair.
The experience lists error patterns, trait-related triggers, and actionable preference guidance transferred to unseen scenarios.}
\label{fig:app_experience_example}
\end{figure}

Figure~\ref{fig:app_experience_example} gives one induced experience example.
}{%
One induced experience example is summarized below.
}
The proposal model does not rewrite the candidate actions directly.
It produces textual guidance such as when the scenario requires proactive greeting and the participant has low extraversion but high robot trust, avoid overselecting assertive approaching actions and prefer verbally initiated, spatially conservative options.
The guidance is model agnostic in form but tuned per participant--model pair because error patterns depend on both the annotator profile and the target model behavior.

\end{document}